\documentclass[sigplan,10pt,dvipsnames]{acmart}
\usepackage[10pt]{moresize}
\usepackage{xspace}
\usepackage{listings}
\usepackage{hyperref}
\usepackage{cleveref}
\usepackage{todonotes}
\usepackage{soul}
\usepackage{float}
\floatstyle{boxed} 
\restylefloat{figure}
\usepackage{FiraMono}
\usepackage[T1]{fontenc}

\newcommand{\ruleName}[1]{\textbf{\footnotesize #1}}
\newcommand{\metaDef}{\mathrel{\mathop:}=}
\newcommand{\cmid}{\textcolor{NavyBlue}{\ \mid\!\!\mid \ }}
\newcommand{\optional}[1]{\textcolor{NavyBlue}{[\!\![} #1 \textcolor{NavyBlue}{]\!\!]}}
\newcommand{\many}[5]{\textcolor{NavyBlue}{\left(\textcolor{black!100}{#1}\right)_{#3 \in \mathcal{\uppercase{#5}}}}}
\newcommand{\some}[5]{\textcolor{NavyBlue}{\left(\textcolor{black!100}{#1}\right)_{#3 \in \mathcal{\uppercase{#5}}}}}
\newcommand{\toSet}[5]{\{{#1} \mid {#3} \in \mathcal{\uppercase{#5}}\}}

\newcommand{\instRel}{\ \preceq \ }

\newcommand{\keyword}[1]{\textcolor{Violet}{\mathtt{#1}}}
\newcommand{\mkeyword}[1]{\textbf{#1}}

\newcommand{\evalRel}{\rightsquigarrow}
\newcommand{\stepRel}{\rightarrowtail}

\newcommand{\elevate}{ELEVATE\xspace}
\newcommand{\code}[1]{\texttt{\footnotesize #1}}

\newcommand{\rise}{RISE\xspace}
\lstdefinestyle{miniElevateStyle}{
    basicstyle=\ttfamily\footnotesize
}

\usepackage{enumitem}

\usepackage{tikz}
\newcommand\tikzmark[1]{\tikz[remember picture,overlay]\node[below] (#1) {};}
\usetikzlibrary{tikzmark}
\usetikzlibrary{calc}

\acmJournal{PACMPL}
\acmVolume{1}
\acmNumber{PLDI} 
\acmArticle{1}
\acmYear{2021}
\acmMonth{1}
\acmDOI{} 
\startPage{1}

\setcopyright{none}

\setcitestyle{numbers,sort&compress}

\makeatletter
\renewcommand\@formatdoi[1]{\ignorespaces}
\makeatother
\renewcommand\footnotetextcopyrightpermission[1]{}

\usepackage{booktabs}   
\usepackage{subcaption} 

\lstdefinelanguage{elevate}{
    morecomment=[l]{--}, 
    morecomment=[l]{//}, 
    morecomment=[s]{/*}{*/}, 
    keywords=[1]{
        let,
        lam,
        rec,
        in,
        match,
        with,
        type,
        as,
        forall
    },
    keywords=[2]{
        Rise,
        Id,
        Name,
        Lam,
        Param,
        Body,
        App,
        Fun,
        Arg,
        Primitive,
        Map,
        Zip,
        Reduce,
        map,
        fun
    },
    keywords=[3]{
        RewriteResult,
        Success,
        Failure,
        Strategy
    },
    keywords=[4]{
        f,
        g,
        h,
        x,
        xs
    },
}
\begin{document}
\thispagestyle{empty}

\title{Row-Polymorphic Types for Strategic Rewriting}


\author{Rongxiao Fu}
\affiliation{
  \institution{University of Glasgow}           
  \country{United Kingdom}                   
}
\email{rongxiao.fu@glasgow.ac.uk}          

\author{Xueying Qin}
\affiliation{
  \institution{The University of Edinburgh}           
  \country{United Kingdom}                   
}
\email{xueying.qin@ed.ac.uk}

\author{Ornela Dardha}
\orcid{nnnn-nnnn-nnnn-nnnn}             
\affiliation{
  \institution{University of Glasgow}           
  \country{United Kingdom}                   
}
\email{ornela.dardha@glasgow.ac.uk}         

\author{Michel Steuwer}
\orcid{0000-0001-5048-0741}             
\affiliation{
  \institution{The University of Edinburgh}           
  \country{United Kingdom}                   
}
\email{michel.steuwer@ed.ac.uk}         
%
%
%
%
\begin{abstract}

We present a type system for \emph{strategy languages} that express program transformations as compositions of rewrite rules.
Our row-polymorphic type system assists compiler engineers to write correct strategies by statically rejecting non meaningful compositions of rewrites that otherwise would fail during rewriting at runtime.
Furthermore, our type system enables reasoning about how rewriting transforms the shape of the computational program.
We present a formalization of our language at its type system and demonstrate its practical use for expressing compiler optimization strategies.

Our type system builds the foundation for many interesting future applications, including verifying the correctness of program transformations and synthesizing program transformations from specifications encoded as types.
\end{abstract}



\maketitle

\section{Introduction}
Rewrite systems find applications in many domains ranging from logic~\cite{DBLP:conf/alp/Marchiori94} and theorem provers~\cite{DBLP:journals/jlp/HsiangKLR92} to program transformations~\cite{VISSER2001109}.
In many domains, it is sufficient to specify a set of \emph{rewrite rules} -- each specifying a small rewrite step -- which are applied (possibly non-deterministically) until a normal form is reached or no rule is applicable anymore.

For practical program transformations this is not adequate.
While \emph{rewrite rules} are a straightforward choice for encoding single program transformation, it might be required to apply a rule only to a subpart of a program, multiple rules in a specific order, or en-/disable rules during a specific phase.
To control the application of rewrite rules, \emph{strategy languages} such as Stratego~\cite{visser1998building,martin2008stratego} have been proposed --
\citeauthor{kirchner2015rewriting}~\cite{kirchner2015rewriting} provides a recent overview of the field.
These strategy languages enable \emph{strategic rewriting} by composing individual rewrite rules into larger rewrite strategies that encode the transformation of an entire program. 

\citeauthor{bastianhagedorn2020achieving}~\cite{bastianhagedorn2020achieving}  describe how the \elevate strategy language is used to encode and control the application of traditional compiler optimizations such as loop-tiling achieving performance comparable to the traditionally designed TVM compiler~\cite{chen2018tvm} for deep learning.
This picks up the trend of increased importance of efficiency in many application domains of today and the future.
For example, the breakthrough success of deep learning has only been possible thanks to carefully optimized software making efficient use of modern parallel hardware.
In the TVM compiler, optimization decisions are encoded in a so-called \emph{schedule} where performance engineers select from a fixed set of exposed compiler transformations to optimize their deep learning application.
In \elevate, strategic rewriting gives developers even greater flexibility as they are free to encode novel program transformations -- possibly domain- or hardware-specific -- as strategies and precisely control their application.

However, developing strategies that encode meaningful program transformations is not easy.
One reason is that current strategy languages provide little to no support for developers for the correct composition of rewrites, for example such as preventing the composition of two rewrites where the first rewrite results in a program shape that the second rewrite cannot be applied to.
On the other hand, statically-typed mainstream programming languages feature types and type systems that guide developers to use composition correctly, for example, by only allowing the composition of functions with appropriately matching types.

This leads us to our research questions:
\emph{How can we design a type systems for strategy languages in order to obtain strong guarantees, statically, for correct composition of rewrites?}

In this paper, we address this question and we propose a type system for a strategy language used for program transformations.
More specifically, we present a row-polymorphic type system for the \elevate strategy language.
We encode the grammar of the abstract syntax tree (AST) of the computational programs that are rewritten as a recursive variant types.
This enables the representation of computational programs at the \elevate type level.
Rewrite rules in \elevate are encoded as functions whose type encapsulates the shape of the program matched by the rule as well as the program shape after applying the rewrite.
Finally, checking that two rewrites compose correctly now simply becomes type checking, namely that their function types compose.

Rewrite rules in \elevate are implemented via pattern matching over the computational program's shape.
When the expected shape is matched, then the rewritten program is returned, otherwise a failure case is returned.
Because pattern matching is such a central part of implementing rewrite rules and strategies, our type system provides exhaustive checking, thus enforcing that all possible cases are covered.
Our practical implementation provides a desugaring mechanism that automatically expands, convenient to write but hard to analyze, \emph{complex} patterns and expand them into easier to check, simple patterns.


To summarize, in this paper we present the following:
\begin{itemize}
    \item \textbf{Row-polymorphic type system}: we present for the first time a row-polymorphic type system for strategy languages that integrates exhaustive checking of pattern matching (\Cref{sec:type-system}).
    
    \item \textbf{Properties of type system}
 Our type system statically guarantees that rewrite strategies (i) will not fail at runtime due to a missing case in pattern matching; (ii) do not contain erroneous access of fields in variants/records and (iii) do not contain a branch in pattern matching that is statically guaranteed not to be reached (discussion at the end of \Cref{sec:type-system}).

    \item \textbf{Implementation of the type system}: we present a practical implementation of our type system including type inference and pattern expansion (\Cref{sec:implementation}), and
    \item \textbf{Program transformation case study}: we present detailed examples and a discussion of the practical use and benefits of the type system for strategic rewriting of program transformations (\Cref{sec:by-example}).
\end{itemize}

\section{Background and Motivation}
\label{sec:background}

\paragraph{Program transformations as rewrite rules}
Rewriting is a convenient method for encoding program transformations, particular in functional programs.
%
%
%
For example, the Glasgow Haskell Compiler (GHC) allows defining rewrite rules in the source code with the \code{RULES} pragma so the compiler can perform transformation based on the given rules \cite{peytonjones2001playing}. This mechanism works as the implementation of deforestation/fusion \cite{gill1993ashortcut} in the standard library and significantly improves its performance. The implementation of a more advanced optimization technique, stream fusion, in the \code{vector} library \cite{coutts2007stream, peytonjones2013exploiting} also heavily uses rewrite rules.

A classic example of rewrite rules is the map fusion rule, and it can be defined in GHC as follows:

\begin{lstlisting}[xleftmargin=.5cm,xrightmargin=.5cm,language={}]
  {-# RULES "mapFusion" forall f g xs.
      map g (map f xs) = map (g . f) xs #-}
\end{lstlisting}

This definition shows the common structure of a rewrite rule: a collection of meta-variables (\code{f}, \code{g} and \code{xs} here), the left-hand side to be matched against the input program and the right-hand side to be instantiated based on the matching result.
The rule describes a transformation replacing the composition of two consecutive list mappings (\code{map f} and \code{map g}) with a single mapping using the composition of the two element processing functions (\code{f . g}), so the intermediate data \code{map g xs} is eliminated, saving the time of storing/reading it into/from the memory. Practically, GHC uses the more general and powerful producer \& consumer model \cite{gill1993ashortcut} for deforestation/fusion, but the major part of the implementation still relies on the \code{RULES} pragma.

However, GHC executes rewriting in a straightforward but inflexible way, where all active rules will be exhaustively applied (if applicable) to the program during AST traversal. To gain some necessary yet limited flexibility, GHC provides a phase control mechanism which allows the users to configure which group of rules to apply at each simplifier phase in the compiling pipeline \cite{peytonjones2001playing}. This is a basic method for organizing the application of rules, but it is not sufficient.

Consider the following expression: \code{map h (map g (map f xs))}.
%
There could be three different rewriting results from this expression by applying the \code{mapFusion} rule: \code{map (h.g) (map f xs)}, \code{map h (map (g.f) xs)}, and \code{map (h.g.f) xs}.
Depending on the context of this expression each of these results might be preferred, e.g., each result could be the starting point of subsequent rewriting steps that produce an overall better optimized program.
GHC produces the last result by applying the \code{mapFusion} rule greedily as many times as possible.
This provides limitations in practice, as the simple GHC rewriting mechanism has limited its capability to perform some complex optimization tasks according to \cite{farmer2014hermit}, where a key transformation in enhanced stream fusion cannot be expressed by the default GHC rewriting system, and a GHC plugin, HERMIT \cite{farmer2015hermit}, is used to solve this problem.

\paragraph{From Rewrite Rules to Strategies}
To overcome the inflexibilities of simple rewrite systems, \citeauthor{visser1998building} proposed \emph{strategic rewriting} by designing a language for composing individual rewrite rules into strategies that precisely control the rewriting process~\cite{visser1998building}.
The Stratego strategy language~\cite{martin2008stratego} has been widely applied in program analysis, transformation and
synthesis.
It also affected many other strategy languages, including a core part of the HERMIT plugin mentioned above, the Kansas University Rewrite Engine (KURE) \cite{sculthorpe2014kure}, which is a strategy language embedded in Haskell, focusing on typed transformations of ASTs.
More recently, it has inspired the \elevate language that has been used to control the application of compiler optimizations encoded as rewrite rules~\cite{bastianhagedorn2020achieving}.

The crucial idea of \elevate (and Stratego) is to encode a rewrite rule as a function with a return type that enables composition in a monadic style.
The map fusion rule from before is expressed in \elevate as\footnote{The syntax here is adjusted from~\cite{bastianhagedorn2020achieving} to match the syntax used in the rest of the paper and to avoid confusion.}:
\begin{lstlisting}[numbers=left,xleftmargin=.5cm,xrightmargin=.25cm]
let mapFusion: Rise -> RewriteResult Rise =
  lam expr
    match expr with <
      @\tikzmark{pBegin}@map g (map f xs)@\tikzmark{pEnd}@ =>
        Success (@\tikzmark{rBegin}@map fun(x => g (f x)) xs@\tikzmark{rEnd}@)
    | _ => Failure
    >
\end{lstlisting}%
\begin{tikzpicture}[remember picture,overlay,transform canvas={yshift=0.2em}]
  \begin{scope}[line width=8pt, color=PineGreen, opacity=0.3]
    \draw (pic cs:pBegin) -- (pic cs:pEnd);
    \draw (pic cs:rBegin) -- (pic cs:rEnd);
  \end{scope}
\end{tikzpicture}%
The green highlighted parts are the computational program that is rewritten expressed in the functional language \rise~\cite{bastianhagedorn2020achieving}.
The \rise program \code{p} is pattern matched and if the pattern in line 3 is matched the rewritten \rise program is returned in line 4 wrapped in \code{Success}.
Otherwise \code{Failure} is returned.
Functions with the type \code{Rise -> RewriteResult Rise} are called \emph{strategies} in \elevate.

Strategy languages provides support for constructing more complex strategies from simpler one.
These strategy combinators~\cite{martin2008stratego} are introduced to compose strategies and control their applications.
Some commonly used strategy combinators include the sequential combinator \code{;} which combines two strategies by applying the first one and passing its output to the second, the left-biased choice combinator \code{<+} which prefers the result from applying its first strategy and only resorts to the second strategy if the first one fails, and the \code{try} combinator which tries to apply a strategy, but will not change the input program in the case of failure.

There are also strategy combinators which specify the way in which an input program is traversed. For example, the \code{topdown} combinator performs a pre-order traverse over the computational program's AST and apply the argument strategy to all encountered sub-terms, while the \code{bottomup} combinator performs a post-order traverse.

\citeauthor{bastianhagedorn2020achieving} discuss in~\cite{bastianhagedorn2020achieving} how \elevate is used in practice to optimize real-word code such as matrix multiplication computations achieving high performance comparable to traditional compilers for deep learning.

\paragraph{Types for Strategy Languages}
It is easy to make mistakes when writing optimization strategies in \elevate.
We know that static types are a good tool to assist the programmer in writing meaningful programs by avoiding errors that can be spotted by the compiler statically without running the program.
Unfortunately, while \elevate (and generally strategy languages) are functional languages with a basic type system, e.g. distinguishing function types from the type of the computational program AST, their type systems are of limited use as all strategies share the same function type.

But there is clearly static information that the type system could capture to assist programs.
For example, it is obvious that the \code{mapFusion} rule will only succeed if the input program is a sequential composition of two \code{map}s.
Furthermore, the sequential composition \code{mapFusion ; mapFusion} would require the input expression to be three consecutive \code{map}s, but it still has same function type as \code{mapFusion}.
Similarly, by design it is obvious that the \code{try} combinator will never fail, but this is not reflected in its type.

In this paper, we propose to use a row-polymorphic type system to capture the change of a program shape by a strategy statically to improve the understanding of programmers and the static reasoning of tools.
%
Row polymorphism has been originally introduced by \citeauthor{remy1989type}~\cite{remy1989type} and \citeauthor{wand1991type}~\cite{wand1991type} and allows to model extensible data types~\cite{morris19row}.
We present a row-polymorphic type system that models polymorphic variant and record types enabling the encoding of the computational program's AST that is rewritten at the \elevate type level.
With this we are able to distinguish the types of individual strategies and provide useful detail in the types assisting the programmer and tools in their statical reasoning.

\section{Typed \elevate by Example}
\label{sec:by-example}
\label{sec:prog}
In this section we give an overview of our type system and its benefits for strategic rewriting. 
The type system is formally introduced in \Cref{sec:type-system} and its implementation discussed in \Cref{sec:implementation}.
Our type system is designed for the \elevate language, however it is not limited to only \elevate and can directly be applied to other strategy languages. In this section we will show examples inspired from using \elevate for optimizing computational programs expressed in the functional language \rise as discussed in~\cite{bastianhagedorn2020achieving}.

We will introduce a formal syntax in \Cref{sec:type-system}.
For this section we will use standard functional programming notation and point out syntactic constructs as we go along.


\subsection{Encoding Representation of Programs in Types}
\label{sec:rise}
To enable the type system to reason about strategies transforming programs we must encode a representation of computational programs as types.

We encode the abstract syntax tree (AST) of the functional computational language \rise as follows:
\begin{lstlisting}[numbers=left,xleftmargin=.5cm]
type Rise = t as <
    Id:  { Name: Nat | * }
  | Lam: { Param: Nat | Body: t | * }
  | App: { Fun: t | Arg: t | * }
  | Primitive: <Map: {*} | Zip: {*} | Reduce: {*} | *>
  |  * >
\end{lstlisting}
This is a recursive row-polymorphic variant type written with angle brackets \code{< >} that enclose a \emph{row} -- a list of label-type-pairs whose elements are separated by a vertical bar \code{|} and terminated by the empty row \code{*}.
The variant type lists all possible syntactic options for a typical functional language:
\lstinline{Id}entifier, \lstinline{Lam}bda expressions, and function \lstinline{App}lication.
Additionally, \rise provides a set of built-in \lstinline{Primitive}s for expressing data-parallel computations.

Record types are written with curly braces \code{\{ \}} enclosing a row where each label corresponds to a record field such as the \lstinline!Name! of an identifier.
Lambdas have two record fields: for the \lstinline{Param}eter and the \lstinline{Body};
and function applications have also two fields: the \lstinline{Fun}ction to call and the \lstinline{Arg}ument.

In the definition, the type variable \code{t} enables to refer to the overall type, matching the recursive definition of an AST.

\rise expressions such as \lstinline{map g (map f xs)} can be desugared into \lstinline{app(app(map, g), app(app(map, f), xs))} and then be represented in \elevate with a matching type as follows:
\begin{lstlisting}
-- expression:
App {Fun: App {Fun: Primitive Map | Arg: g} |
     Arg: App {Fun: App {Fun: Primitive Map | Arg: f} |
               Arg: xs}}
-- type:
: <App: {Fun: <App: {
               Fun: <Primitive: <Map: {*} | > | > |
               Arg: g | *} | > |
         Arg: <App: {
               Fun: <App: {
                     Fun: <Primitive: <Map: {*} | > | > |
                     Arg: f | *} | > |
               Arg: xs | *} | > | } | >
\end{lstlisting}

%
\subsection{Strategies in Typed \elevate}
\label{sec:strategies-in-elevate}
Strategies are encoded as functions. In the original \elevate a strategy has the type:
\lstinline{Rise -> RewriteResult Rise}.
In our row-polymorphic typed \elevate, strategies have a more precise type allowing to capture the shape of the input program \lstinline!p1! and the rewritten program \lstinline!p2!:
\begin{lstlisting}
type Strategy = forall p1 p2. p1 -> RewriteResult p2
\end{lstlisting}
where \lstinline!RewriteResult! is the rewritten program or failure:
\begin{lstlisting}
type RewriteResult = forall p.
    < Success: p | Failure: {*} | * >
\end{lstlisting}

\paragraph{\code{MapFusion} in typed \elevate}
\Cref{lst:typed-mapFusion} shows the \\\code{mapFusion} rewrite rule that we have seen earlier implemented in our row-polymorphicly typed \elevate.

\begin{lstlisting}[float=t, label={lst:typed-mapFusion}, caption={Implementation of the \code{mapFusion} rewrite rule as a strategy in typed \elevate.}, numbers=left,xleftmargin=.65cm,mathescape]
let mapFusion = lam expr = match expr with <
  -- map g (map f xs)
  App {Fun: App {Fun: Primitive Map | Arg: g} |
       Arg: App {Fun: App {Fun: Primitive Map | 
                           Arg: f} |
                 Arg: xs}} =>
  -- Success ( map fun(x => g (f x)) xs )
  Success (App {Fun: App {Fun: Primitive Map |
    Arg: Lam { Param: 0 | Body: App {Fun: g |
      Arg: App {Fun: f | Arg: Id {Name: 0}}}}} | 
    Arg: xs})>
\end{lstlisting}

\begin{lstlisting}[float=b, label={lst:mapFusion-type}, caption={Inferred type for the \code{mapFusion} strategy.}, numbers=left,xleftmargin=.65cm, mathescape]
  <App: {Fun: <App: {
          Fun: <Primitive: <Map: {*} | *> | *> |
          Arg: g | } | *> |
         Arg: <App: {
          Fun: <App: {
            Fun: <Primitive: <Map: {*} | *> | *> |
            Arg: f | } | *> |
          Arg: xs | } | *> | } | *>
->
  <Success: <App: {
    Fun: <App: {
      Fun: <Primitive: <Map: {*} | > | > |
      Arg: <Lam: {Param: <0: {*} | > | Body: <App: {
        Fun: g |
        Arg: <App: {
          Fun: f |
          Arg: <Id: {Name: <0: {*} | > | *} | > |*
            } | > | * } | > | *} | > | * } | > |
      Arg: xs | * } | > | >
\end{lstlisting}

The green highlighted parts in line 3--5 are the pattern and in lines 7--9 the rewritten expression.
To encode the binding in the lambda expression we use de Bruijn indices~\cite{Bruijn1972LambdaCN}.

This implementation is similar to the implementation in the original weakly typed \elevate language with one important difference:
before we had to provide a default case for when the input program did not have the expected shape;
now we rely on the more advanced type system to ensure that this strategy is only applicable when the input program has the expected shape.
This is also reflected by the inferred type in \Cref{lst:mapFusion-type} capturing the behavior of the strategy statically.
With syntactic sugar we might read this type as:
\begin{lstlisting}[xleftmargin=.5cm,xrightmargin=.5cm]
< map g (map f xs) | * > ->
  < Success (map fun(x => g (f x)) xs) | >
\end{lstlisting}
The argument type in line 1--8 describes the shape of the input program corresponding to the pattern from \Cref{lst:typed-mapFusion}.
The return type describes the rewritten program shape in lines 10---19.
The variables \lstinline!g!, \lstinline!f!, and \lstinline!xs! in the pattern of \Cref{lst:typed-mapFusion} correspond directly to the type variables in \Cref{lst:mapFusion-type}.

\paragraph{Compatibility of types}
A detailed look at the return type reveals that not all rows in variant types are terminated by the empty row \code{*}.
For a simpler example, we consider the type \lstinline!<Failure: {*} | >! we infer for  \lstinline!Failure!.
For cases such as this when the final entry in a row is omitted this indicates the presence of an implicit type variable, so that the type is equivalent to \lstinline!<Failure: {*} | t>!.
This implicit type variable makes the type compatible with other types.
For example, we want to be able to pass the \lstinline!Failure! value to a function that expects a \lstinline!RewriteResult! as argument.
To make the argument and parameter type compatible we simply instantiate the implicit type variable with the \lstinline!Success: p | *! row, which would not be possible without the implicit type variable.
Instantiation of row variables that ensures the well formedness of rows is formalized in \Cref{sec:type-system}.

\paragraph{Exhaustive Checking and Pattern Elaboration}
Pattern matching is a central aspect of our language as it is integral for implementing strategies such as seen in \Cref{lst:typed-mapFusion}.
An important guarantee that we want to provide is that a strategy does not fail at runtime due to a missing case in the pattern matching.
We also want to warn developers if they provide a case for which we can statically conclude that it is impossible to be reached.

\Cref{sec:type-system} details the typing rules that provide these guarantees.
To simplify our formal system we reduce the complexity of patterns that we need to reason about.
More specifically, we only consider \emph{simple} patterns that are either a variable pattern (\lstinline!x!) or a label $l$ followed by a variable (\lstinline[mathescape]!$l$ x!).
But the pattern in \Cref{lst:typed-mapFusion} is clearly neither of these two simple cases.
How do we deal with such more \emph{complex} patterns?

\Cref{sec:implementation} discusses a pattern elaboration mechanism that rewrites complex patterns using into a sequence of simple patterns that our type system is able to check for exhaustiveness.
\Cref{lst:mapFusion-pattern-expanded} shows the implementation of the \lstinline{mapFusion} strategy after pattern elaboration.
Variables introduced by the pattern elaboration are prefixed with \lstinline!#!.
The single \lstinline!match! expression has been replaced with a series of nested \lstinline!match! expressions, each with a simple label or variable pattern.
The record pattern in \Cref{lst:typed-mapFusion} is now decomposed into individual accesses to the record fields.
The typing rules presented in \Cref{sec:type-system} guarantee that this expression will not fail at runtime due to a pattern matching failure.

\begin{lstlisting}[float=t, label={lst:mapFusion-pattern-expanded}, caption={\code{mapFusion} with elaborated pattern matching.}, numbers=left,xleftmargin=.65cm]
let mapFusion = lam expr = match expr with <
  App #x1 => match #x1.Fun with <
    App #x2 => match #x2.Fun with <
      Primitive #x3 => match #x3 with <
        Map => match #x1.Arg with <
          App #x4 => match #x4.Fun with <
            App #x5 => match #x5.Fun with <
              Primitive #x6 => match #x6 with <
                Map => match #x2.Arg with <
                  g => match #x5.Arg with <
                    f => match #x4.Arg with <
                      xs => Success (...) >>>>>>>>>>>
\end{lstlisting}


\subsection{Strategy Combinators in Typed \elevate}
\label{sec:generalProg}
Strategy languages provide combinators for composing simple strategies into more complicated strategies.
\Cref{lst:combinators} shows the implementation of some key combinators and their inferred row-polymorphic types in \elevate.

The \lstinline!id! and \lstinline!fail! strategies are not combinators but useful building blocks for larger compositions.
The inferred types are as expected reflecting at the type level the certain behavior of the strategy.

The \lstinline!seq!uential combinator first applies strategy \lstinline!fs! (line 10) before inspecting the result and applying the second strategy \lstinline!ss! in line 11 only if the first strategy has been successful.
The inferred type reflects this behavior:
the successful rewritten program \lstinline!p2! in the type of the first strategy (line 7) must be the input program of the second strategy (line 8).
The composed strategy has a type (line 9) combining the input type of the first strategy (\lstinline!p1!) with the return type of the second.
Here the \emph{row variable} \lstinline!r! can either represent the \lstinline!Success! case with the rewritten program, or it can be instantiated with the empty row type when the second strategy is guaranteed to result in \lstinline!Failure! (as the \lstinline!fail! strategy does).
The notation here indicates that the row variable \lstinline!r! can be instantiated with any label except \lstinline!Failure!.
The kinding system for row variables will be explained in detail in \Cref{sec:type-system}.

The \lstinline!lChoice! combinator prefers the successful outcome of the first strategy and only applies the second strategy if the first fails.
The inferred type is interesting, as it seems that the two strategies as well as the combined strategy all return the same rewritten program \lstinline!p2! on success (lines 15--17).
This is because it is not statically possible to determine which successfully rewritten program will be chosen at runtime and, therefore, the type system unifies both possible outcomes.

Finally, the \lstinline!try! combinator applies the given strategy and in the failure case reverts back to unchanged input program.
The implementation uses \lstinline!lChoice! and \lstinline!id! and we can observe from the inferred type that this combinator can never fail due to the absence of \lstinline!Failure! in the return type.
This type can be easily derived from the type of \lstinline!lChoice! with the row variable \lstinline!r! instantiated with the empty row as \lstinline!id! can never result in \lstinline!Failure!.

\begin{lstlisting}[float=t, label={lst:combinators}, caption={Strategy combinators and their types in \elevate.}, numbers=left,xleftmargin=.65cm]
let id: p -> <Success: p | > =
  lam expr = Success expr

let fail: p -> <Failure: {*} | > =
  lam expr = Failure

let seq: (p1 -> <Success: p2 | Failure: {*} | *>) ->
         (p2 -> <Failure: {*} | r: ~{Failure}>) ->
         (p1 -> <Failure: {*} | r: ~{Failure}>) =
  lam fs = lam ss = 
    lam expr1 = match (fs expr1) with <
        Success expr2 => ss expr2
      | Failure => Failure
    >

let lChoice: 
  (p1 -> <Success: p2 | Failure: {*} | *>) ->
  (p1 -> <Success: p2 | r: ~{Success}>) ->
  (p1 -> <Success: p2 | r: ~{Success}>) =
    lam fs = lam ss = 
      lam expr1 = match (fs expr1) with <
          Success expr2 => Success expr2
        | Failure => ss expr1
      >

let try: (p1 -> <Success: p1 | Failure: {*} | *>) ->
         (p1 -> <Success: p1 | >) =
  lam s = lChoice s id
\end{lstlisting}

\subsection{Safe Compositions in Typed \elevate}
Using the combinators we can now compose strategies.
For example, sequentially composing the \lstinline!mapFusion! strategy twice:
\lstinline!mapFusion ; mapFusion! results in the following type that we show only using syntactic sugar for brevity:
\begin{lstlisting}[xleftmargin=.5cm,xrightmargin=.5cm]
< map h (map g (map f xs)) | * > ->
  < map fun(x => fun(y => h (g y)) (f x)) xs |
    Failure: {*} | >
\end{lstlisting}
From the type we observe that the input program of this program transformation must be a sequence of three applications of the \lstinline!map! function and that the rewritten program contains only a single instance of \lstinline!map!.

But what happens when we try to compose strategies that are impossible to compose?
For example, consider the \lstinline!reduceMapFusion! rewrite rules from~\cite{bastianhagedorn2020achieving}:
\begin{lstlisting}[xleftmargin=.5cm,xrightmargin=.5cm, language={}]
reduce g init (map f xs) =
  reduce fun(a => fun(x => g a (f x))) init xs
\end{lstlisting}
Encoding this rule as an \elevate strategy and trying to sequentially compose this rule with the \lstinline!mapFusion! strategy leads to a type error.
This is because the type system cannot find an instantiation of the type and row variables that would make the two types compatible and the composition safe.

We have seen in this section the practical use of an advanced type system for a strategy language encoding program transformations.
Our type system provides important guarantees ensuring the safe composition of strategies as well as checking the exhaustiveness of pattern matching guaranteeing that pattern matching cannot fail at runtime.
Next, we give a formal account of our type system.

\section{Typed \elevate, Formalized}
\label{sec:type-system}
In this section, we present a core calculus for \elevate. We show the syntax of terms and types and the type system leveraging row-polymorphism.

For the formalization in this section, unless otherwise stated, elements in \(\textcolor{NavyBlue}{NavyBlue}\) are meta-level descriptions, for e.g., the square brackets \(\optional{\ }\) indicate \emph{option} in the EBNF grammar;  the indexed multiple occurrences (possibly separated by either \(\ \mid\ \) or \(\ ,\ \)) of a syntactical construct \(expr\) are collectively represented by \(\many{expr_i}{,}{i}{0}{n}\), where the index (written as \(i\), \(j\), \(k\), \(p\) or \(q\)) ranges over a possibly empty subset (written as \(\mathcal{M}\), \(\mathcal{N}\), \(\mathcal{U}\) or \(\mathcal{V}\)) of the set of natural numbers.

\subsection{Syntax}
\label{sec:syntax}
  \begin{figure}
  \footnotesize
  \begin{alignat*}{4}
  \mathrm{Terms} \quad
  &e \ &\metaDef \quad & x \cmid e_1 \ e_2 \cmid \lambda \ x = e \cmid \\
  &&&\keyword{let} \ f = e_1 \ \keyword{in} \ e_2 \cmid \keyword{fix} \cmid \\
  &&&l \ e \cmid \{\many{l_i: e_i}{\mid}{i}{0}{n}\} \cmid e.l \cmid e.-l \cmid \\
  &&&e.\optional{+}\{\many{l_i : e_i}{\mid}{i}{0}{n}\} \cmid \\
  &&&\keyword{match} \ e \ \keyword{with} \ \langle\optional{\{\cdot\} \Rightarrow e_1}\rangle \cmid \\
  &&&\keyword{match} \ e \ \keyword{with} \ \langle l \ x_1 \Rightarrow e_1 \mid x_2 \Rightarrow e_2 \rangle
\end{alignat*}
\caption{Syntax of terms}
\label{fig:ast}
\end{figure}

\paragraph{Terms}
Figure \ref{fig:ast} shows the syntax of terms and patterns.
Terms (denoted by $e$) include common constructs such as variables ranged over $x,y,z$, term applications, lambda abstractions, let-bindings, and the fixed point combinator. In addition, terms include the following new constructs: label applications (denoted by \(l \ e\)) for constructing variant values, record constructors (denoted by \(\{\many{l_i: e_i}{\mid}{i}{0}{n}\}\)), field accesses (denoted by \(e.l\)), field removals (denoted by \(e.-l\)), record modifications (denoted by \(e.\{\many{l_i : e_i}{\mid}{i}{0}{n}\}\)), record extensions (denoted by \(e.+\{\many{l_i : e_i}{\mid}{i}{0}{n}\}\)), where the order of label-term pairs is insignificant and labels are all different; and finally pattern matchings: an \elevate term $e$ can be matched with the empty pattern (\(\keyword{match} \ e \ \keyword{with} \ \langle\rangle\)), the unit (empty record) pattern (\(\keyword{match} \ e \ \keyword{with} \ \langle\{\cdot\} \Rightarrow e_1\rangle\)) or the variant pattern (\(\keyword{match} \ e \ \keyword{with} \ \langle l \ x_1 \Rightarrow e_1 \mid x_2 \Rightarrow e_2 \rangle\)) which introduces a variable $x_1$ for the case of label $l$, and a variable $x_2$ representing the rest of the cases. In the rest of this paper, we may omit the $x_1$ for simplicity if it is immediately matched against the empty record, that is, \(\keyword{match} \ e \ \keyword{with} \ \langle l \Rightarrow e_1 \mid x_2 \Rightarrow e_2 \rangle\) means \(\keyword{match} \ e \ \keyword{with} \ \langle l \ x_1 \Rightarrow \keyword{match} \ x_1 \ \keyword{with} \ \langle\{\cdot\} \Rightarrow e_1\rangle \mid x_2 \Rightarrow e_2 \rangle\)

\begin{figure}
  \footnotesize
  \begin{alignat*}{4}
  \mathrm{Kinds} \quad 
   &\kappa \ &\metaDef \quad &\mathcal{T} \cmid \mathcal{R}\\
   &\mathcal{R} \ &\metaDef \quad &L \cmid \neg L\\
   &L \ &\metaDef \quad &\{\many{l_i}{,}{i}{0}{n}\}\\
  \\[-.5em]
  \mathrm{Types} \quad 
  &t \ &\metaDef \quad &\alpha \cmid \tau \cmid \alpha \ \keyword{as} \ \tau \\
  &\tau \ &\metaDef \quad &t_1 \to t_2 \cmid \{ \rho \} \cmid \langle \rho \rangle\\
  &\rho \ &\metaDef \quad &\alpha \cmid \cdot \cmid l: t \mid \rho\\
  \\[-.5em]
  \mathrm{Schemes} \quad 
  &\sigma \ &\metaDef \quad &t \cmid \forall \ (\alpha: \kappa).\ \sigma
\end{alignat*}
\caption{Syntax of kinds and types}
\label{fig:types}
\end{figure}

\paragraph{Kinds and types}
Figure \ref{fig:types} shows the syntax of kinds and types in \elevate, whose design is influenced by Blume et al. \cite{blume06cases}.
Types are of two kinds: \(\mathcal{T}\)--for ordinary types, and \(\mathcal{R}\)--for row types. Ordinary types (denoted by \(t\)) include type variables (denoted by \(\alpha\)), type constructor applications or \textit{contractive} types \cite{macQueen84ideal} (denoted by \(\tau\)), and equi-recursive types (denoted by \(\alpha \ \keyword{as} \ \tau\)). We require types appearing under an equi-recursive binder to be contractive, which excludes meaningless types such as \(\alpha \ \keyword{as} \ \alpha\) and guarantees the existence of an unique solution to the recursive equation(s) \cite{remy2013type, im13contractive}. Contractive types (denoted by \(\tau\)) include \textbf{function types} (denoted by \(t_1 \to t_2\)), \textbf{record types} (denoted by \(\{ \rho \}\)) and \textbf{variant types} (denoted by \(\langle \rho \rangle\)). 

Row types (denoted by \(\rho\)) are sequences of label-type pairs $l: t$ ending with row variables or empty rows (denoted by \(\cdot\)), where the order of the label-term pairs is insignificant, and labels are all distinct. Rows are differentiated from ordinary types by their kinds, row kinds (denoted by \(\mathcal{R}\)), which can be positive (denoted by \(L\), a finite subset of the set of all labels) or negative (denoted by \(\neg L\), a cofinite subset of the set of all labels) descriptions of sets of labels. The set of all possible labels and the set of all possible variable names are disjoint. Unlike \cite{blume06cases} where a row kind is associated with a row variable and describe the finite set of labels that the row variable \textbf{must not} contain, kind of a row in \elevate represents the possibly infinite/cofinite set of labels which \textbf{can} appear in this row. For e.g., the kind of the empty row is \(\{\}\), indicating that no label can appear in the empty row; given a row variable \(r\) of kind \(\neg \{A, B\}\), the row \((A: a \mid r)\) has kind \(\neg \{B\}\), which means any label except \(B\) can appear in the row (cf. detailed explanation in Section \ref{sec:type}).

Type schemes (denoted by \(\sigma\)) represent possibly universally quantified types. The kind of the bound type variable should be specified at the binding site.

\subsection{Type System}
\label{sec:type}

Before the type system for \elevate, we give in Figure~\ref{fig:env} the kinding (denoted by \(\Delta\)) and typing environments (denoted by \(\Gamma\)): they can be either empty (denoted by \(\cdot\)) or extended with a type variable and its kind (denoted by \(\alpha: \kappa\)) or a variable and its type scheme (denoted by \(x: \sigma\)), respectively.

\begin{figure}
  \footnotesize
  \vspace{-.5em}
  \begin{alignat*}{4}
    \mathrm{Kinding \ Environment} \quad 
    &\Delta \ &\metaDef &\cdot \cmid \Delta, \alpha: \kappa\\
    \\[-1em]
    \mathrm{Typing \ Environment} \quad 
    &\Gamma \ &\metaDef &\cdot \cmid \Gamma, x: \sigma
    \vspace{-.5em}
    \end{alignat*}
  \caption{Syntax of kinding and typing environments}
  \label{fig:env}
  \end{figure}

As mentioned in Section \ref{sec:syntax}, the kind of a row represents the set of labels that can appear in this row, and the set can be described positively or negatively. When only negative row kinds, i.e., the set of labels that the row \textbf{must not} contain, are used, this kinding mechanism is similar to the lack relation in \cite{blume06cases} or the row kind in \cite{pottier04ML}, disallowing the construction of ill-formed rows like \((A: \alpha \mid A: \alpha \mid r)\). With positive row kinds, more meaningful restrictions can be added to rows.

For e.g., given a row variable \(r\) of kind \(\neg \{True\}\), meaning (the substitution for) \(r\) must not contain the label \(True\), the type \(\langle True: \{\cdot\} \mid r \rangle\) can be unified with the whimsical type \(\langle True: \{\cdot\} \mid Apple: \{\cdot\} \mid \cdot \rangle\) because the kind of \((Apple: \{\cdot\} \mid \cdot)\) is \(\{Apple\}\), meaning the label \(Apple\) can appear in this row, and it is compatible with \(\neg \{True\}\). However, if the kind \(\{False\}\) is initially assigned to \(r\), the unification above is impossible and \(\langle True: \{\cdot\} \mid r \rangle\) can only be unified with more sensible types like \(\langle True: \{\cdot\} \mid False: \{\cdot\} \mid \cdot \rangle\). Although this kinding mechanism does not stop users from getting seemingly useless types like \(\langle True: \{\cdot\} \mid False: \langle\cdot\rangle \mid \cdot \rangle\), it allows more precise specification of the ranges of labels than solely negative row kinds.

Figure \ref{fig:row} gives the computation rules for row kinds, defining the row kind subset relation (denoted by \(\sqsubseteq\)) and the row kind extension operator (denoted by \(+\)). The row kind subset relation is essentially defined as the subset relation of the sets represented by the row kinds, and the row kind extension operator is defined as element insertion without duplicates. 

Figure \ref{fig:kinding} gives the kinding rules, which also work as the well-formedness rules for types. Kinding judgments are of the form \(\Delta \vdash t: \kappa\), stating that the type \(t\) has kind \(\kappa\) (and is well formed) in the kinding environment \(\Delta\). Most of the kinding rules are straightforward. The most important rule among them is \(\ruleName{K-RowExtension}\) showing the usage of the row kind extension operator: it extends the kind with a label and rules out rows containing repeated labels.

\begin{figure}
\footnotesize
\begin{alignat*}{1}
&\frac{
  \toSet{l_i}{,}{i}{0}{n} \subseteq \toSet{l_j}{,}{j}{0}{m}
}{
  \{\many{l_i}{,}{i}{0}{n}\} \sqsubseteq \{\many{l_j}{,}{j}{0}{m}\}
} \ \ruleName{R-Subset-pp}
\\\\[-.5em]
&\frac{
  \toSet{l_j}{,}{j}{0}{m} \subseteq \toSet{l_i}{,}{i}{0}{n}
}{
  \neg \{\many{l_i}{,}{i}{0}{n}\} \sqsubseteq \neg \{\many{l_j}{,}{j}{0}{m}\}
} \ \ruleName{R-Subset-nn}
\\\\[-.5em]
&\frac{
  \toSet{l_i}{,}{i}{0}{n} \cap \toSet{l_j}{,}{j}{0}{m} = \emptyset
}{
  \{\many{l_i}{,}{i}{0}{n}\} \sqsubseteq \neg \{\many{l_j}{,}{j}{0}{m}\}
} \ \ruleName{R-Subset-pn}
\\\\[-.5em]
&\frac{
  \begin{aligned}
  &l \notin \toSet{l_i}{,}{i}{0}{n} \\
  &\{l\} \cup \toSet{l_i}{,}{i}{0}{n} = L
  \end{aligned}
}{
  \{\many{l_i}{,}{i}{0}{n}\} + l = L
} \ \ruleName{R-Ext-p}
\\\\[-.5em]
&\frac{
  \begin{aligned}
  &l \in \toSet{l_i}{,}{i}{0}{n} \\
  &\toSet{l_i}{,}{i}{0}{n} \setminus \{l\} = L
  \end{aligned}
}{
  \neg \{\many{l_i}{,}{i}{0}{n}\} + l = \neg L
} \ \ruleName{R-Ext-l}
\end{alignat*}
\caption{Computation rules for row kinds}
\label{fig:row}
\end{figure}

\begin{figure}
\footnotesize
\begin{alignat*}{1}
&\frac{\alpha: \kappa \in \Delta}{\Delta \vdash \alpha: \kappa} \ \ruleName{K-TVar}
\\\\[-.5em]
&\frac{\Delta \vdash t_1: \mathcal{T} \quad \Delta \vdash t_2: \mathcal{T}}{\Delta \vdash t_1 \to t_2: \mathcal{T}} \ \ruleName{K-FunctionType}
\\\\[-.5em]
&\frac{\Delta \vdash \rho: \mathcal{R}}{\Delta \vdash \{\rho\}: \mathcal{T}} \ \ruleName{K-RecordType}
\\\\[-.5em]
&\frac{\Delta \vdash \rho: \mathcal{R}}{\Delta \vdash \langle \rho \rangle: \mathcal{T}} \ \ruleName{K-VariantType}
\\\\[-.5em]
&\frac{\Delta, \alpha: \mathcal{T} \vdash \tau: \mathcal{T}}{\Delta \vdash \alpha \ \keyword{as} \ \tau: \mathcal{T}} \ \ruleName{K-RecursiveType}
\\\\[-.5em]
&\frac{}{\Delta \vdash \cdot: \{\}} \ \ruleName{K-EmptyRow}
\\\\[-.5em]
&\frac{\Delta \vdash \rho: \mathcal{R} \quad \Delta \vdash t: \mathcal{T} \quad \mathcal{R} + l = \mathcal{R}^{ext}}{\Delta \vdash (l: t \mid \rho): \mathcal{R}^{ext}} \ \ruleName{K-RowExtension}
\end{alignat*}
\caption{Kinding rules}
\label{fig:kinding}
\end{figure}

From \Cref{fig:typing} to \ref{fig:match}, we presents the typing rules for terms in \elevate, with typing judgments of the form \(\Delta;\Gamma \vdash e: t\), stating that the term \(e\) has type \(t\) in the kinding environment \(\Delta\) and typing environment \(\Gamma\), and we always assume but omit for simplicity that \(t\) and all the types in \(\Gamma\) are well-kinds in \(\Delta\). The notation \(ftv(t)\) or \(ftv(\Gamma)\) stand for the set of free type variables in type \(t\) or typing environment \(\Gamma\), respectively. The typing rules for the lambda-calculus subset of \elevate (\(\ruleName{T-Var}\),\(\ruleName{T-App}\) and \(\ruleName{T-Lam}\)) are standard. The type scheme instantiation relation \(\instRel\) used by the \(\ruleName{T-Var}\) rule is defined in Figure \ref{fig:inst}, where substituting type \(t\) for type variable \(\alpha\) in type scheme \(\sigma\) is denoted as \(\sigma[\alpha \mapsto t]\). As shown by \(\ruleName{T-Inst-Row}\), a universally quantified row variable can only be instantiated by a row whose kind is the subset of the variable, equivalently speaking, by a \textit{"not more general"} row. This rule ensures that well-formed rows in a type scheme are still well-formed after instantiation.

The typing rules for let-binding (\(\ruleName{T-Let}\)) and the fixed point combinator (\(\ruleName{T-Fix}\)) are also standard. We use the strict version of the fixed point combinator in \elevate. 

The rule \(\ruleName{T-Label}\) types a variant value with a label \(l\) and an expression \(e\). It can be considered as an inlined instantiation of the type scheme \(\forall \ (r: \neg\{l\}). \ \langle l: t \mid r \rangle\), hence the row kind \(\mathcal{R}\) is the subset of \(\neg \{l\}\).

\begin{figure}
\footnotesize
\begin{alignat*}{1}
&\frac{
  \begin{aligned}
  &x: \sigma \in \Gamma \quad \Delta \vdash \sigma \instRel t
  \end{aligned}
}{\Delta;\Gamma \vdash x: t} \ \ruleName{T-Var}
\\\\[-.5em]
&\frac{\Delta;\Gamma \vdash f: t_1 \to t_2 \quad \Delta;\Gamma \vdash e: t_1}{\Delta;\Gamma \vdash f \ e: t_2} \ \ruleName{T-App}
\\\\[-.5em]
&\frac{
  \Delta;\Gamma, x: t_x \vdash e: t_t
}{
\Delta;\Gamma \vdash \lambda \ x = e: t_x \to t_e} \ \ruleName{T-Lam}
\\\\[-.5em]
&\frac{
  \begin{aligned}
    &\some{\alpha_i}{}{i}{0}{n} = ftv(t_1) \setminus ftv(\Gamma) \\
    &\Delta, \some{\alpha_i: \kappa_i}{}{i}{0}{n};\Gamma \vdash e_1: t_1 \\
    &\sigma = \some{\forall \ (\alpha_i: \kappa_i).}{}{i}{0}{n}t_1 \quad \Delta;\Gamma, f: \sigma \vdash e_2: t_2
  \end{aligned}
}{
  \Delta;\Gamma \vdash \keyword{let} \ f = e_1 \ \keyword{in} \ e_2: t_2
} \ \ruleName{T-Let}
\\\\[-.5em]
&\frac{}{
  \Delta;\Gamma \vdash \keyword{fix} : ((t_1 \to t_2) \to t_1 \to t_2) \to t_1 \to t_2
} \ \ruleName{T-Fix}
\\\\[-.5em]
&\frac{\Delta;\Gamma \vdash e: t \quad \Delta \vdash \rho: \mathcal{R} \quad \mathcal{R} \sqsubseteq \neg \{l\}}{\Delta;\Gamma \vdash l \ e: \langle l: t \mid \rho\rangle} \ \ruleName{T-Label}
\end{alignat*}
\caption{Typing rules for basic terms}
\label{fig:typing}
\end{figure}

  \begin{figure}
  \footnotesize
  \begin{alignat*}{1}
  &\frac{
    \Delta \vdash t: \mathcal{T} \quad \alpha \notin ftv(t)
  }{
    \Delta \vdash \forall \ (\alpha: \mathcal{T}).\ \sigma \instRel \sigma[\alpha \mapsto t]
  } \ \ruleName{T-Inst-Type}
  \\\\[-.5em]
  &\frac{
  \Delta \vdash \rho: \mathcal{R}^{inst} \quad \mathcal{R}^{inst} \sqsubseteq \mathcal{R} \quad \alpha \notin ftv(\rho)
  }{
    \Delta \vdash \forall \ (\alpha: \mathcal{R}).\ \sigma \instRel \sigma[\alpha \mapsto \rho]
  } \ \ruleName{T-Inst-Row}
  \\\\[-.5em]
  &\frac{
  \Delta \vdash \sigma_0 \instRel \sigma_1 \quad \Delta \vdash \sigma_1 \instRel \sigma_2
  }{
    \Delta \vdash \sigma_0 \instRel \sigma_2
  } \ \ruleName{T-Inst-Trans}
  \end{alignat*}
  \caption{Type scheme instantiation rules}
  \label{fig:inst}
  \end{figure}

Figure \ref{fig:rec} gives the standard rolling and unrolling rules for equi-recursive types.
Figure \ref{fig:recOps} is a collection of the typing rules for record operations. Unlike the type of the newly created variant value in rule \(\ruleName{T-Label}\), it is impossible to make the underlying row in the type of a newly created record contain more labels than the presented ones. Thus, \(\ruleName{T-RecordCons}\) states that the type of the record ends with an empty row, which is the least general row. The rest of the rules are straightforward: field(s) should exist in the record to be modified, accessed, or deleted, and to extend a record, the to-be-added field(s) should not previously exist in the record.

\begin{figure}
  \footnotesize
  \begin{alignat*}{1}
  &\frac{\Delta;\Gamma \vdash e : \alpha \ \keyword{as} \ \tau}{\Delta;\Gamma \vdash e : \tau[\alpha \mapsto \alpha \ \keyword{as} \ \tau]} \ \ruleName{T-Unroll}
  \\\\[-.5em]
  &\frac{\Delta;\Gamma \vdash e : \tau[\alpha \mapsto \alpha \ \keyword{as} \ \tau]}{\Delta;\Gamma \vdash e : \alpha \ \keyword{as} \ \tau} \ \ruleName{T-Roll}
  \end{alignat*}
  \caption{Typing rules for equi-recursive types}
  \label{fig:rec}
\end{figure}

\begin{figure}
\footnotesize
\begin{alignat*}{1}
&\frac{
\begin{aligned}
&\forall \ i \in \mathcal{N}, \Delta;\Gamma \vdash e_i: t_i
\end{aligned}
}{\Delta;\Gamma \vdash \{\many{l_i: e_i}{\mid}{i}{0}{n}\} : \{\many{l_i: t_i}{\mid}{i}{0}{n} \mid \cdot\}} \ \ruleName{T-RecordCons}
\\\\[-.5em]
&\frac{
\begin{aligned}
&\Delta;\Gamma \vdash e: \{\many{l_i: t_i}{\mid}{i}{0}{n} \mid \rho\} \\
&\forall \ i \in \mathcal{N}, \Delta;\Gamma \vdash e_i: t_i
\end{aligned}
}{
  \Delta;\Gamma \vdash e.\{\many{l_i: e_i}{\mid}{i}{0}{n}\} : \{\many{l_i: t_i}{\mid}{i}{0}{n} \mid \rho\}
} \ \ruleName{T-RecordMod}
\\\\[-.5em]
&\frac{
\begin{aligned}
&\Delta;\Gamma \vdash e: \{\rho\} \quad \Delta \vdash \rho: \mathcal{R} \quad \mathcal{R} \sqsubseteq \neg \{\many{l_i}{,}{i}{0}{n}\}\\
&\forall \ i \in \mathcal{N}, \Delta;\Gamma \vdash e_i: t_i
\end{aligned}
}{
  \Delta;\Gamma \vdash e.+\{\many{l_i: e_i}{\mid}{i}{0}{n}\} : \{\many{l_i: t_i}{\mid}{i}{0}{n} \mid \rho\}
} \ \ruleName{T-RecordExt}
\\\\[-.5em]
&\frac{\Delta;\Gamma \vdash e: \{l: t \mid \rho\}}{\Delta;\Gamma \vdash e.l: t} \ \ruleName{T-FieldAccess}
\\\\[-.5em]
&\frac{\Delta;\Gamma \vdash e: \{l: t \mid \rho\}}{\Delta;\Gamma \vdash e.-l: \{\rho\}} \ \ruleName{T-FieldDel}
\end{alignat*}
\caption{Typing rules for record operations}
\label{fig:recOps}
\end{figure}

Pattern matching is an essential part of \elevate, and the corresponding typing rules are given in Figure \ref{fig:match}. \(\ruleName{T-Void}\) and \(\ruleName{T-Unit}\) are defined in the standard way. The premises of \(\ruleName{T-Match}\) generalize the type of the matched expression \(e\), split the generalized type into two parts, and assign them to \(x_1\) and \(x_2\), respectively. Generalization during pattern matching is not common, but this behavior can be found in OCaml and \cite{castagna2016set}, and it allows more programs to typecheck. In other words, in \elevate, pattern matching can be used as an "advanced" form of let-binding, which simultaneously introduces polymorphic variables and analyses different cases of an expression.

\begin{figure}
\footnotesize
\begin{alignat*}{1}
&\frac{
  \Delta;\Gamma \vdash e: \langle\cdot\rangle
}{
  \Delta;\Gamma \vdash \keyword{match} \ e \ \keyword{with} \ \langle\rangle : t_{rhs}
} \ \ruleName{T-Void}
\\\\[-.5em]
&\frac{
  \Delta;\Gamma \vdash e: \{\cdot\} \quad \Delta;\Gamma \vdash rhs: t_{rhs}
}{
  \Delta;\Gamma \vdash \keyword{match} \ e \ \keyword{with} \ \langle \{\cdot\} \Rightarrow rhs \rangle : t_{rhs}
} \ \ruleName{T-Unit}
\\\\[-.5em]
&\frac{
\begin{aligned}
  &\some{\alpha_i: \kappa_i}{}{i}{0}{n} = ftv(t) \setminus ftv(\Gamma) \\
  &\some{\alpha_j: \kappa_j}{}{j}{0}{m} = ftv(\langle\rho\rangle) \setminus ftv(\Gamma) \\
  &\Delta, \some{\alpha_i: \kappa_i}{}{i}{0}{n} \cup \some{\alpha_j: \kappa_j}{}{j}{0}{m};\Gamma \vdash e: \langle l: t \mid \rho\rangle \\
  &\sigma_{x1} = \some{\forall \ (\alpha_i: \kappa_i).}{}{i}{0}{n}t \quad \sigma_{x2} = \some{\forall \ (\alpha_j: \kappa_j).}{}{j}{0}{m}\langle\rho\rangle \\
  &\Delta;\Gamma, x_1: \sigma_{x1} \vdash rhs_1: t_{rhs} \quad \Delta;\Gamma, x_2: \sigma_{x2} \vdash rhs_2: t_{rhs}
\end{aligned}
}{
\begin{aligned}
  &\Delta;\Gamma \vdash \keyword{match} \ e \ \keyword{with} \ \langle
  l \ x_1 \Rightarrow rhs_1 \mid x_2 \Rightarrow rhs_2
  \rangle : t_{rhs}
\end{aligned}
} \ \ruleName{T-Match}
\end{alignat*}
\caption{Typing rules for pattern matching}
\label{fig:match}
\end{figure}

\subsection{Operational Semantics}
The operational semantics of \elevate is given by Figure \ref{fig:value} and Figure \ref{fig:eval} following the style used in \cite{wright94soundness}. Figure \ref{fig:value} provides the definitions of \elevate values and the evaluation contexts. Figure \ref{fig:eval} provides the reduction rules (named with the prefix \ruleName{ST-}) and the stepping relation (denoted by \(\stepRel\)) for the small-step operational semantics of \elevate. All the rules are straightforward.

\begin{figure}
  \footnotesize
  \begin{alignat*}{4}
    \mathrm{Values} \quad 
    &v \ &\metaDef \ \ & l \ v \cmid \{\many{l_i: v_i}{\mid}{i}{0}{n}\} \cmid \lambda \ x = e \cmid \keyword{fix} \\
    &&&\mathrm{Evaluation \ Contexts} \\
    \quad 
    &E \ &\metaDef \ \ & [] \cmid E \ e \cmid v \ E \cmid \\
    &&&\keyword{let} \ f = E \ \keyword{in} \ e \cmid \\
    &&&l \ E \cmid \{\many{l_i: v_i}{\mid}{i}{0}{n} \mid l: E \mid \many{l_j: e_j}{\mid}{j}{0}{m}\} \cmid \\
    &&&E.l \cmid E.-l \cmid \\
    &&&E.\optional{+}\{\many{l_i : e_i}{\mid}{i}{0}{n}\} \cmid \\
    &&&v.\optional{+}\{\many{l_i: v_i}{\mid}{i}{0}{n} \mid l: E \mid \many{l_j: e_j}{\mid}{j}{0}{m}\} \cmid \\
    &&&\keyword{match} \ E \ \keyword{with} \ \langle\optional{\{\cdot\} \Rightarrow e}\rangle \cmid \\
    &&&\keyword{match} \ E \ \keyword{with} \ \langle l \ x_1 \Rightarrow e_1 \mid x_2 \Rightarrow e_2\rangle
  \end{alignat*}
  \caption{Values in \elevate}
  \label{fig:value}
\end{figure}

\begin{figure}
  \footnotesize
  \begin{alignat*}{1}
  &(\lambda \ x = e) \ v \evalRel e[x \mapsto v] \quad \ruleName{ST-App}
  \\\\[-.5em]
  &\keyword{let} \ f = v \ \keyword{in} \ e_2 \evalRel e_2[f \mapsto v] \quad \ruleName{ST-Let}
  \\\\[-.5em]
  &\keyword{fix} \ v \evalRel v \ (\lambda \ x = \keyword{fix} \ v \ x) \quad \ruleName{ST-Fix}
  \\\\[-.5em]
  &\{\many{l_i: v_i}{\mid}{i}{0}{m} \mid l: v \mid \many{l_j: v_j}{\mid}{j}{0}{n}\}.l \evalRel v \quad \ruleName{ST-FieldAccess}
  \\\\[-.5em]
  &\begin{aligned}
      &\{\many{l_i: v_i}{\mid}{i}{0}{m} \mid l: v \mid \many{l_j: v_j}{\mid}{j}{0}{n}\}.-l \\
      &\qquad \evalRel \{\many{l_i: v_i}{\mid}{i}{0}{m} \mid \many{l_j: v_j}{\mid}{j}{0}{n}\}
    \end{aligned} \quad \ruleName{ST-FieldDel}
  \\\\[-.5em]
  &\begin{aligned}
      &\{\many{l_i: v_i}{\mid}{i}{0}{m} \mid \many{l_j: v_j}{\mid}{j}{0}{n}\}.\{\many{l_i: v'_i}{\mid}{i}{0}{m}\} \\ 
      &\qquad \evalRel \{\many{l_i: v'_i}{\mid}{i}{0}{m} \mid \many{l_j: v_j}{\mid}{j}{0}{n}\}
    \end{aligned}
  \quad \ruleName{ST-RecordMod}
  \\\\[-.5em]
  &\begin{aligned}
      &\{\many{l_i: v_i}{\mid}{i}{0}{m}\}.+\{\many{l_j: v_j}{\mid}{j}{0}{n}\} \\ 
      &\qquad \evalRel \{\many{l_i: v_i}{\mid}{i}{0}{m} \mid \many{l_j: v_j}{\mid}{j}{0}{n}\}
    \end{aligned} \quad \ruleName{ST-RecordExt}
  \\\\[-.5em]
  &\keyword{match} \ \{\} \ \keyword{with} \ \langle \{\cdot\} \Rightarrow e \rangle \evalRel e \quad \ruleName{ST-Match-Unit}
  \\\\[-.5em]
  &\begin{aligned}
    &\keyword{match} \ l \ v \ \keyword{with} \ \langle l \ x_1 \Rightarrow e_1 \mid x_2 \Rightarrow e_2\rangle \\
    &\qquad \evalRel e_1[x_1 \mapsto v] 
    \end{aligned} \quad \ruleName{ST-Match-Match}
  \\\\[-.5em]
  &\begin{aligned}
    &\keyword{match} \ l \ v \ \keyword{with} \ \langle l' \ x_1 \Rightarrow e_1 \mid x_2 \Rightarrow e_2\rangle \\
    &\qquad \evalRel e_2[x_2 \mapsto l \ v] 
   \end{aligned} \quad \ruleName{ST-Match-Skip}
  \\\\[-.5em]
  &E[e_1] \stepRel E[e_2] \ \textit{iff} \ e_1 \evalRel e_2 \quad \ruleName{S-Context}
  \end{alignat*}
  \caption{Small-step operational semantics of \elevate}
  \label{fig:eval}
\end{figure}

\subsection{Properties of the row-polymorphic type system}
\label{sec:properties}

\begin{lemma}[Subject Reduction]
  \label{lem:preservation}
  If $\Delta;\cdot \vdash e_1: t$, and $e_1 \evalRel e_2$, then $\Delta;\cdot \vdash e_2: t$.
\end{lemma}

\begin{lemma}[Progress]
  \label{lem:progress}
  If $\Delta;\cdot \vdash e: t$, then either $e$ is a value, or there exists an $\hat{e}$ such that $e \stepRel \hat{e}$.
\end{lemma}

\begin{theorem}[Type Soundness]
  If $\Delta;\cdot \vdash e: t$, then either $e$ is a value, or there exists an $\hat{e}$ such that $e \stepRel \hat{e}$ and $\Delta;\cdot \vdash \hat{e}: t$.
\end{theorem}

The proofs can be found in Appendix \ref{app:proof}.

To be more specific, this type system guarantees that strategies that type check:
\begin{itemize}[leftmargin=5mm]
\item do not fail at runtime due to a missing case in pattern matching (guaranteed by the type soundness theorem); and

\item do not access a non-existent field in a record (guaranteed by the type soundness theorem); and

\item do not contain records or variants involving fields or cases tagged by the same label (guaranteed by the kinding and well-formedness rules); and

\item do not contain a \emph{dead branch} in pattern matching that is statically guaranteed not to be reached.
\end{itemize}

\paragraph{On the detection of dead branches}
By only relying on the type system we are able to detect two forms of dead code:

\begin{itemize}[leftmargin=5mm]

  \item A dead branch is detected when a label used in the pattern has already been matched by previous branches or cannot occur in the type of the matched expression.
  For example:
  \begin{itemize}
    \item when the same label $A$ is repeated multiple times:\\
    \(\strut\quad \keyword{match} \ x \ \keyword{with} \ \langle A \Rightarrow rhs_0\)\\
    \(\strut\quad\quad \mid y \Rightarrow \keyword{match} \ y \ \keyword{with} \ \langle A \Rightarrow rhs_1 \mid \dots \rangle\rangle\)
    \item when \(x\) with type \(\langle B: \{\cdot\} \mid r\rangle\) and where the kind of \(r\) is \(\neg \{A, B\}\) is matched against label $A$:\\
    \(\strut\quad \keyword{match} \ x \ \keyword{with} \ \langle A \Rightarrow rhs \mid \dots \rangle\)
  \end{itemize}
  
  \item A dead branch is detected when a row variable is exploited that is only there for type compatibility purposes as discussed earlier in \Cref{sec:by-example}.
        Formally, these kind of row variables do not occur free in the typing context.
        They do not contain meaningful information (for pattern matching) and can always be substituted by the empty row.
  %
  %
  Examples ruled out by this include:
  \begin{itemize}

    \item the inferred type for \(True\) is \(\langle True: \{\cdot\} \mid r\rangle\) where the kind of \(r\) is \(\neg \{True\}\) but \(r\) does not occur free in the typing context.
    This disallows matching the value \(True\) against the pattern \(False\):\\
    \(\strut\quad\keyword{match} \ True \ \keyword{with} \ \langle False \Rightarrow rhs \mid \dots \rangle\)
			
    \item similarly, we disallow matching the remainder \(x\) after the \(True\) case against any other types except the empty variant. The variable \(x\) has type \(\langle r \rangle\) but \(r\) does not occur free in the typing context and is substituted with the empty row, ruling out expressions in the following form where the expected type of \(x\) in \({rhs}_1\) is not \(\langle\cdot\rangle\):\\
    \(\strut\quad\keyword{match} \ True \ \keyword{with} \ \langle True \Rightarrow rhs_0 \mid x \Rightarrow rhs_1 \rangle\)
  \end{itemize}

\end{itemize}

There are other forms of dead branches that our current type system is unable to detect.
For example, the inferred type of \((seq \ \textit{fail} \ \textit{id})\) is \(t_0 \to \langle Failure: \{\cdot\} \mid Success: t_1 \mid r \rangle\), but we know that the result can never be \(Success\).
We have to deal with the \(Success\) case when we analyze the result with pattern matching.

However, the type actually tell us that the \(Success\) case is unnecessary because \(t_1\) does not occur free in the typing context and it can be substituted by the empty type -- since there is only one way to use the empty type, a branch for the case is useless.
This particular situation, is similar to the second form of dead code detection mentioned above.
The underlying common idea is that, if a type \(t\) contains type variables which do not occur free in the typing context, and substituting these type variables with empty type/row will make \(t\) isomorphic to the empty type, then there is no need to deal with a value of this type in pattern matching.
In the general case, allowing the types which are isomorphic to the empty type to be used as empty type require more complex type system and put forward challenges for type inference.
A possible solution to this is to use semantic subtyping as in \cite{castagna2016set}, which gives a complete and sound algorithm to solve type constraints with semantic subtyping, in the price of losing principal solutions.

\section{Implementation}
\label{sec:implementation}

After introducing the type system formally in \Cref{sec:type-system} we now discuss its practical implementation focusing on two important aspects: pattern elaboration and type inference.

\subsection{Pattern Elaboration}
\label{sec:elab}
As mentioned in Section \ref{sec:strategies-in-elevate} and \ref{sec:type}, pattern matching is an important part of \elevate, but patterns that are easy to write for programmers may be too \textit{complex} for type inference and exhaustiveness checking. Thus, pattern elaboration is used to bridge this gap between the formalized typed \elevate and its practical implementation. In this section, unless otherwise specified, the term "pattern" only refers to the simple patterns defined in Figure \ref{fig:ast}.

Figure \ref{fig:patSyn} shows the abstract syntax of complex patterns (denoted by \(\tilde{\pi}\)). In comparison with the syntax of patterns in Figure \ref{fig:ast}, complex patterns allow the recursive occurrence of complex patterns inside a label (denoted by \(l \ \tilde{\pi}\)) and the usage of record patterns (denoted by \(\{\many{l_i: \tilde{\pi}_i}{\mid}{i}{0}{n}\}\)), which match the distinct fields \(\many{l_i}{\mid}{i}{0}{n}\) respectively with complex patterns \(\many{\tilde{\pi}_i}{\mid}{i}{0}{n}\). As in the formalization, the order of the fields is insignificant in a record pattern. Linearity checks will be performed to make sure that each variable only appears once in a complex pattern.

\begin{figure}
  \footnotesize
  \begin{alignat*}{4}
  \mathrm{Complex \ Patterns} \quad
  &\tilde{\pi} \ &\metaDef \quad &x \cmid l \ \optional{\tilde{\pi}} \cmid \{\many{l_i: \tilde{\pi}_i}{\mid}{i}{0}{n}\}
  \\
  \mathrm{Field \ Access \ Forms} \quad 
  &\delta \ &\metaDef \quad &x\optional{.l}\optional{.\{\}}
  \\
  \mathrm{Match \ IDs} \quad 
  &\ell \ &\metaDef \quad &\mathbb{N} \cmid \ell \mid \mathbb{N}
  \\
  \mathrm{Match \ Chains} \quad 
  &\varpi \ &\metaDef \quad &e^{\ell} \cmid \keyword{match}^{\ell} \ \delta \ \keyword{with} \ \langle\some{\pi_i \Rightarrow \varpi_i}{\mid}{i}{0}{n}\rangle
  \end{alignat*}
  \caption{Syntax of complex patterns, field access forms, match IDs and match chains}
  \label{fig:patSyn}
\end{figure}

For a single pattern matching expression with more than one branch, the basic idea of pattern elaboration is to convert the complex pattern in each branch into nested pattern matching expressions only using (simple) patterns, and then merging all branches to generate a decision tree \cite{maranget08decision}. Figure \ref{fig:patElab} shows the pseudo-code of pattern elaboration. Functions \(patExpn\) and \(merge\) perform the conversion and merging mentioned above, respectively. Function \(foldl1\) performs left-folding of list and takes the first element of the list as the starting value. Function \(sort\) rearranges the order of the nested pattern matching expressions generated by \(patExpn\) to get a more efficient result. Since the efficiency of pattern matching is not of major concern in this work, the \(sort\) function will not be discussed in details here. Finally, function \(refine\) adjusts expressions in the decision tree to get a more precise type inference result and \(desugar\) convert match chains into ordinary pattern matching expressions.

\begin{figure}
  \footnotesize
  \vspace{-1em}
  \begin{alignat*}{1}
    &patElab(\keyword{match} \ x \ \keyword{with} \ \langle\many{\tilde{\pi}_i \Rightarrow rhs_i}{\mid}{i}{0}{n}\rangle) \\
    &\quad = desugar(refine(x \mapsto x, \ foldl1(merge, \\
    &\qquad \many{sort(patExpn(\ell^e_i, \ 0, \ u, \ \keyword{match} \ x \ \keyword{with} \ \langle\tilde{\pi}_i \Rightarrow rhs_i\rangle))}{\mid}{i}{0}{n}))
  \end{alignat*}
  \vspace{-1.25em}
  \caption{The pseudo-code of pattern elaboration}
  \label{fig:patElab}
\end{figure}

\begin{figure*}
\footnotesize
\begin{alignat*}{3}
  \; \qquad \qquad && patExpn(\ell^e, \ &\ell, \ u, \ \keyword{match} \ \delta \ \keyword{with} \ \langle x \Rightarrow e \rangle) = \keyword{match}^{\ell} \ \delta \ \keyword{with} \ \langle x \Rightarrow e^{\ell^e} \rangle \\
  \; \qquad \qquad && patExpn(\ell^e, \ &\ell, \ u, \ \keyword{match} \ \delta \ \keyword{with} \ \langle l \Rightarrow e \rangle) = \keyword{match}^{\ell} \ \delta \ \keyword{with} \ \langle l \Rightarrow e^{\ell^e} \rangle \\
  \; \qquad \qquad && patExpn(\ell^e, \ &\ell, \ u, \ \keyword{match} \ \delta \ \keyword{with} \ \langle l \ x \Rightarrow e \rangle) = \keyword{match}^{\ell} \ \delta \ \keyword{with} \ \langle l \ x \Rightarrow e^{\ell^e} \rangle \\
  \; \qquad \qquad && patExpn(\ell^e, \ &\ell, \ u, \ \keyword{match} \ \delta \ \keyword{with} \ \langle l \ \tilde{\pi} \Rightarrow e \rangle) = \keyword{match}^{\ell} \ \delta \ \keyword{with} \ \langle l \ x \Rightarrow \varpi \rangle \\
  \; \qquad \qquad && \quad \mkeyword{where} \quad &x \ \ \mkeyword{is \ fresh} \\
  \; \qquad \qquad && \quad \quad &\varpi = patExpn(\ell^e, \ (\ell \mid 0), \ x, \ \keyword{match} \ x \ \keyword{with} \ \langle \tilde{\pi} \Rightarrow e \rangle) \\
  \; \qquad \qquad && patExpn(\ell^e, \ &\ell, \ u, \ \keyword{match} \ \delta \ \keyword{with} \ \langle \{\} \Rightarrow e \rangle) = \keyword{match}^{\ell} \ \delta.\{\} \ \keyword{with} \ \langle x \Rightarrow e^{\ell^e} \rangle \\
  \; \qquad \qquad && \quad \mkeyword{where} \quad &x \ \ \mkeyword{is \ fresh} \\
  \; \qquad \qquad && patExpn(\ell^e, \ &\ell, \ x, \ \keyword{match} \ x \ \keyword{with} \ \langle \{l:\tilde{\pi}\} \Rightarrow e \rangle) = patExpn(\ell^e, \ \ell, \ x, \ \keyword{match} \ x.l \ \keyword{with} \ \langle \tilde{\pi} \Rightarrow e \rangle) \\
  \; \qquad \qquad && patExpn(\ell^e, \ &(\ell \mid n), \ x, \ \keyword{match} \ x \ \keyword{with} \ \langle \{l:\tilde{\pi} \mid \some{l_i: \tilde{\pi}_i}{\mid}{i}{0}{n}\} \Rightarrow e \rangle) = patExpn(\ell^e, \ (\ell \mid n), \ x, \ \keyword{match} \ x.l \ \keyword{with} \ \langle \tilde{\pi} \Rightarrow v \rangle)[v \mapsto \varpi] \\
  \; \qquad \qquad && \quad \mkeyword{where} \quad &v \ \ \mkeyword{is \ fresh} \\
  \; \qquad \qquad && \quad \quad &\varpi = patExpn(\ell^e, \ (\ell \mid n + 1), \ x, \ \keyword{match} \ x \ \keyword{with} \ \langle \{\some{l_i: \tilde{\pi}_i}{\mid}{i}{0}{n}\} \Rightarrow e \rangle) \\
  \; \qquad \qquad && patExpn(\ell^e, \ &\ell, \ u, \ \keyword{match} \ \delta \ \keyword{with} \ \langle \{\some{l_i: \tilde{\pi}_i}{\mid}{i}{0}{n}\} \Rightarrow e \rangle) = \keyword{match}^{\ell} \ \delta.\{\} \ \keyword{with} \ \langle x \Rightarrow \varpi \rangle \\
  \; \qquad \qquad && \quad \mkeyword{where} \quad &x \ \ \mkeyword{is \ fresh} \\
  \; \qquad \qquad && \quad \quad &\varpi = patExpn(\ell^e, \ (\ell \mid 0), \ x, \ \keyword{match} \ x \ \keyword{with} \ \langle \{\some{l_i: \tilde{\pi}_i}{\mid}{i}{0}{n}\} \Rightarrow e \rangle)
\end{alignat*}
\caption{The pseudo-code of pattern expansion}
\label{fig:patExpn}
\end{figure*}

\begin{figure*}
  \footnotesize
  \begin{alignat*}{1}
    \quad &merge(\keyword{match}^{\ell_a} \ \delta_a \ \keyword{with} \ \langle\many{\pi_i \Rightarrow \varpi_i}{\mid}{i}{0}{n} \mid x_a \Rightarrow \varpi_a\rangle, \ \keyword{match}^{\ell_b} \ \delta_b \ \keyword{with} \ \langle x_b \Rightarrow \varpi_b \rangle) \\
    \quad &\quad \mkeyword{when} \quad length(\ell_a) == length(\ell_b) \ \mkeyword{and} \ \delta_a \simeq \delta_b = \keyword{match}^{\ell_a} \ \delta_b \ \keyword{with} \ \langle\many{\pi_i \Rightarrow merge(\varpi_i, \ \varpi_b[x_b \mapsto \pi_i])}{\mid}{i}{0}{n} \mid x_a \Rightarrow merge(\varpi_a, \ \varpi_b[x_b \mapsto x_a])\rangle \\
    \quad &merge(\keyword{match}^{\ell_a} \ \delta_a \ \keyword{with} \ \langle\many{\pi_i \Rightarrow \varpi_i}{\mid}{i}{0}{n}\rangle, \ \keyword{match}^{\ell_b} \ \delta_b \ \keyword{with} \ \langle x_b \Rightarrow \varpi_b \rangle) \\
    \quad &\quad \mkeyword{when} \quad length(\ell_a) == length(\ell_b) \ \mkeyword{and} \ \delta_a \simeq \delta_b = \keyword{match}^{\ell_a} \ \delta_b \ \keyword{with} \ \langle\many{\pi_i \Rightarrow merge(\varpi_i, \ \varpi_b[x_b \mapsto \pi_i])}{\mid}{i}{0}{n} \mid x_b \Rightarrow \varpi_b\rangle \\
    \quad &merge(\keyword{match}^{\ell_a} \ \delta_a \ \keyword{with} \ \langle\many{\pi_i \Rightarrow \varpi_i}{\mid}{i}{0}{n} \mid l \ x_a \Rightarrow \varpi_a \mid \many{\pi_j \Rightarrow \varpi_j}{\mid}{j}{0}{m}\rangle, \ \keyword{match}^{\ell_b} \ \delta_b \ \keyword{with} \ \langle l \ x_b \Rightarrow \varpi_b \rangle) \\
    \quad &\quad \mkeyword{when} \quad length(\ell_a) == length(\ell_b) \ \mkeyword{and} \ \delta_a \simeq \delta_b = \keyword{match}^{\ell_a} \ \delta_b \ \keyword{with} \ \langle\many{\pi_i \Rightarrow \varpi_i}{\mid}{i}{0}{n} \mid l \ x_a \Rightarrow merge(\varpi_a, \ \varpi_b[x_b \mapsto x_a]) \mid \many{\pi_j \Rightarrow \varpi_j}{\mid}{j}{0}{m}\rangle \\
    \quad &merge(\keyword{match}^{\ell_a} \ \delta_a \ \keyword{with} \ \langle\many{\pi_i \Rightarrow \varpi_i}{\mid}{i}{0}{n} \mid l \Rightarrow \varpi_a \mid \many{\pi_j \Rightarrow \varpi_j}{\mid}{j}{0}{m}\rangle, \ \keyword{match}^{\ell_b} \ \delta_b \ \keyword{with} \ \langle l \Rightarrow \varpi_b \rangle) \\
    \quad &\quad \mkeyword{when} \quad length(\ell_a) == length(\ell_b) \ \mkeyword{and} \ \delta_a \simeq \delta_b = \keyword{match}^{\ell_a} \ \delta_b \ \keyword{with} \ \langle\many{\pi_i \Rightarrow \varpi_i}{\mid}{i}{0}{n} \mid l \Rightarrow merge(\varpi_a, \ \varpi_b) \mid \many{\pi_j \Rightarrow \varpi_j}{\mid}{j}{0}{m}\rangle \\
    \quad &merge(\keyword{match}^{\ell_a} \ \delta_a \ \keyword{with} \ \langle\many{\pi_i \Rightarrow \varpi_i}{\mid}{i}{0}{n} \mid x_a \Rightarrow \varpi_a\rangle, \ \keyword{match}^{\ell_b} \ \delta_b \ \keyword{with} \ \langle \pi_b \Rightarrow \varpi_b \rangle) \\
    \quad &\quad \mkeyword{when} \quad length(\ell_a) == length(\ell_b) \ \mkeyword{and} \ \delta_a \simeq \delta_b = \keyword{match}^{\ell_a} \ \delta_a \ \keyword{with} \ \langle\many{\pi_i \Rightarrow \varpi_i}{\mid}{i}{0}{n} \mid x_a \Rightarrow merge(\varpi_a, \ \keyword{match}^{\ell_b} \ x_a \ \keyword{with} \ \langle \pi_b \Rightarrow \varpi_b \rangle)\rangle \\
    \quad &merge(\keyword{match}^{\ell_a} \ \delta_a \ \keyword{with} \ \langle\many{\pi_i \Rightarrow \varpi_i}{\mid}{i}{0}{n}\rangle, \ \keyword{match}^{\ell_b} \ \delta_b \ \keyword{with} \ \langle \pi_b \Rightarrow \varpi_b \rangle) \\
    \quad &\quad \mkeyword{when} \quad length(\ell_a) == length(\ell_b) \ \mkeyword{and} \ \delta_a \simeq \delta_b = \keyword{match}^{\ell_a} \ \delta_b \ \keyword{with} \ \langle\many{\pi_i \Rightarrow \varpi_i}{\mid}{i}{0}{n} \mid x_b \Rightarrow \varpi_b\rangle \\
    \quad &merge(\keyword{match}^{\ell_a} \ \delta_a \ \keyword{with} \ \langle\many{\pi_i \Rightarrow \varpi_i}{\mid}{i}{0}{n}\rangle, \ \keyword{match}^{\ell_b} \ \delta_b \ \keyword{with} \ \langle \pi_b \Rightarrow \varpi_b \rangle) \\
    \quad &\quad \mkeyword{when} \quad length(\ell_a) == length(\ell_b) = merge(\keyword{match}^{\ell_a} \ \delta_a \ \keyword{with} \ \langle\many{\pi_i \Rightarrow \varpi_i}{\mid}{i}{0}{n}\rangle, \ \keyword{match}^{\ell_a} \ \delta_a \ \keyword{with} \ \langle x \Rightarrow \keyword{match}^{\ell_b} \ \delta_b \ \keyword{with} \ \langle \pi_b \Rightarrow \varpi_b \rangle \rangle) \\
    \quad &\quad \mkeyword{where} \quad x \ \ \mkeyword{is \ fresh} \\
    \quad &merge(\keyword{match}^{\ell_a} \ \delta_a \ \keyword{with} \ \langle\many{\pi_i \Rightarrow \varpi_i}{\mid}{i}{0}{n}\rangle, \ \varpi_b) = merge(\keyword{match}^{\ell_a} \ \delta_a \ \keyword{with} \ \langle\many{\pi_i \Rightarrow \varpi_i}{\mid}{i}{0}{n}\rangle, \ \keyword{match}^{\ell_a} \ \delta_a \ \keyword{with} \ \langle x \Rightarrow \varpi_b \rangle) \\
    \quad &\quad \mkeyword{where} \quad x \ \ \mkeyword{is \ fresh} \\
    \quad &merge(e^{\ell^e}, \varpi_b) = e^{\ell^e}
\end{alignat*}
\caption{The pseudo-code of match chain merging}
\label{fig:mcMerge}
\end{figure*}

\begin{figure*}
\footnotesize
\begin{alignat*}{1}
  \qquad \qquad \qquad \qquad \qquad &refine(S, \ \keyword{match}^{\ell} \ \delta \ \keyword{with} \ \langle\many{\pi_i \Rightarrow \varpi_i}{\mid}{i}{0}{n}\rangle) = \keyword{match}^{\ell} \ \delta \ \keyword{with} \ \langle\many{refineStep(\delta, \ \pi_i \Rightarrow \varpi_i)}{\mid}{i}{0}{n}\rangle\\
  \qquad \qquad \qquad \qquad \qquad &\quad \mkeyword{where} \quad refineStep(x, \ \pi \Rightarrow \varpi) = (\pi \Rightarrow refine((x \mapsto \pi) \circ S, \ \varpi_i)) \\
  \qquad \qquad \qquad \qquad \qquad &\qquad \qquad \; \ \ refineStep(x.\{\}, \ \pi \Rightarrow \varpi) = (\pi \Rightarrow refine((x \mapsto \pi) \circ S, \ \varpi_i)) \\
  \qquad \qquad \qquad \qquad \qquad &\qquad \qquad \; \ \ refineStep(x.l, \ \pi \Rightarrow \varpi) = (\pi \Rightarrow refine((x \mapsto x.-l.+\{l: \pi\}) \circ S, \ \varpi_i)) \\
  \qquad \qquad \qquad \qquad \qquad &\qquad \qquad \; \ \ refineStep(x.l.\{\}, \ \pi \Rightarrow \varpi) = (\pi \Rightarrow refine((x \mapsto x.-l.+\{l: \pi\}) \circ S, \ \varpi_i)) \\
  \qquad \qquad \qquad \qquad \qquad &refine(S, \ e^{\ell^e}) = e^{\ell^e}[S]
\end{alignat*}
\caption{The pseudo-code of match chain refining}
\label{fig:mcRefine}
\end{figure*}

\begin{figure*}
  \footnotesize
  \begin{alignat*}{1}
    \qquad \qquad \qquad \qquad
    &desugar(\keyword{match}^{\ell} \ \delta \ \keyword{with} \ \langle l \Rightarrow \varpi \mid \many{\pi_i \Rightarrow \varpi_i}{\mid}{i}{0}{n}\rangle)\\
    &\quad = \keyword{match} \ \delta \ \keyword{with} \ \langle l \ x \Rightarrow \keyword{match} \ x \ \keyword{with} \ \langle \{\} \Rightarrow desugar(\varpi) \rangle \mid r \Rightarrow desugar(\keyword{match}^{\ell} \ r \ \keyword{with} \ \langle \many{\pi_i \Rightarrow \varpi_i}{\mid}{i}{0}{n}\rangle)\rangle\\
    &\quad \mkeyword{where} \quad x \ \mkeyword{and} \ r \ \ \mkeyword{are \ fresh} \\
    &desugar(\keyword{match}^{\ell} \ \delta \ \keyword{with} \ \langle l\ x \Rightarrow \varpi \mid \many{\pi_i \Rightarrow \varpi_i}{\mid}{i}{0}{n}\rangle)\\
    &\quad = \keyword{match} \ \delta \ \keyword{with} \ \langle l \ x \Rightarrow desugar(\varpi) \rangle \mid r \Rightarrow desugar(\keyword{match}^{\ell} \ r \ \keyword{with} \ \langle \many{\pi_i \Rightarrow \varpi_i}{\mid}{i}{0}{n}\rangle)\rangle\\
    &\quad \mkeyword{where} \quad r \ \ \mkeyword{is \ fresh} \\
    &desugar(\keyword{match}^{\ell} \ \delta \ \keyword{with} \ \langle x \Rightarrow \varpi \rangle) = \keyword{let} \ x = \delta \ \keyword{in} \ desugar(\varpi)\\
    &desugar(\keyword{match}^{\ell} \ \delta \ \keyword{with} \ \langle\rangle) = \keyword{match} \ \delta \ \keyword{with} \ \langle\rangle\\
    &desugar(e^{\ell^e}) = e^{\ell^e}
  \end{alignat*}
  \caption{The pseudo-code of match chain desugaring}
  \label{fig:mcDesugar}
\end{figure*}

Figure \ref{fig:patExpn} shows the pseudo-code of \(patExpn\), meaning "pattern expansion". Specifically, a complex pattern will expand into a series of simple pattern matching expressions. A key part of this process is the conversion of record patterns. Since a record pattern is just multiple complex patterns put together, where each one of them is associated with a label, a straightforward method is replacing a record pattern with a variable pattern \(x\), then selecting each matched field of \(x\), and then recursively performing the conversion for each field. Similar conversion can also be applied to complex label patterns (\(l \ \tilde{\pi}\)). This method naturally covers a special case of the record patterns, namely \(\{\}\): it does not select any field to match, so it matches all records and is directly replaced by a variable pattern during the conversion. An awkward consequence is that a variable pattern does not only match records, but also matches other type of values, so there must be some other way to distinguish an ordinary variable pattern and \(\{\}\). This leads us to the design of the field access forms shown in Figure \ref{fig:patSyn}. A field access form (denoted by \(\delta\)) can be a variable (\(x\)) or a field access (\(x.l\)), optionally followed by an empty record modification (\(x.\{\}\) or \(x.l.\{\}\)). As its name suggests, the field access form expresses the intermediate field accesses required to convert record patterns. Since the empty record modification does nothing but enforce the modified value to be a record, it helps distinguish ordinary variable patterns and variable patterns which used to be record patterns.

With the conversion method above, each single branch of a pattern matching expression can be converted to a biased decision tree, and we use match chains to encode this. Figure \ref{fig:patSyn} shows the definition of match chains. A match chain (denoted by \(\varpi\)) is a multi-way tree-like structure used throughout the pattern elaboration process. It can be either an ordinary expression, i. e., the RHS expression, or a pattern matching expression whose RHS expressions are match chains. Since there can be multiple complex patterns in one record pattern, and a complex pattern can be deeply nested or even repeatedly occurs inside itself, to identify the depths and locations of match chain nodes, a match ID is assigned to each of them. A match ID (denoted by \(\ell\)) is simply a non-empty sequence of natural numbers whose length indicates the depth in a nested complex pattern and the exact numbers indicates the locations in record patterns.

Figure \ref{fig:mcMerge} shows the pseudo-code of match chain merging. With match IDs and field access forms, \(merge\) can easily tell if two match chains are matching the same value (hence can be merged). It should be noted that the second and some other lines of \(merge\) duplicate \(\varpi_b\), while the final line of \(merge\) removes \(\varpi_b\). This is related with another usage of the match IDs: they are also the unique identifiers for RHS expressions. All the duplications and removals of RHS expressions are traced, and if all the occurrences of a RHS expression are removed, a \code{\textcolor{red}{redundant patterns}} error will be reported.

Figure \ref{fig:mcRefine} shows the pseudo-code of match chain refining. The idea behind this definition is straightforward. If a field access form \(\delta\) is matched by a pattern \(\pi\), we know that in the corresponding RHS expression, the actual value of \(\delta\) can only be the expression counterpart of \(\pi\). To get a more precise type inference result, \(refine\) substitutes the identifier in \(\delta\) with \(\pi\) or the corresponding record operations.

Finally, Figure \ref{fig:mcDesugar} shows the pseudo-code of match chain desugaring, which recursively converts each branch in a match chain into cascaded pattern matching expressions.

\subsection{Type Inference}
\label{sec:infer}
The type inference of \elevate follows the widely used Hindley-Milner style~\cite{DBLP:journals/jcss/Milner78} and extends it with row polymorphism.
The implementation of the core part of this inference algorithm, a union-find based unifier supporting equi-recursive types and rows, is largely based on the Huet's unification algorithm \cite{knight1989unification, huet76Alg} and the unifier implementation in the Mini inference engine \cite{pottier04rowkind}. Currently, \elevate does not put any other restrictions besides contractiveness on the form of equi-recursive types.

Compared with the typing rules presented in Section \ref{sec:type}, the type inference algorithm of \elevate is more restrictive for pattern matching expressions. As discussed in Section \ref{sec:properties}, pattern matching branches may become dead code if some special row variables are substituted by empty rows. This form of dead code will be removed during type inference, even if they can have valid type derivations using the rules for type checking. The RHS tracing mechanism mentioned in Section \ref{sec:elab} still works here, and it will report errors when the occurrence count of a RHS expression is reduced to zero, otherwise the branch will be removed silently.

Using our practical implementation, we have implemented all examples from \cite{bastianhagedorn2020achieving} in typed \elevate observing the expected types by running type inference on them and confirming that the presented strategies are well typed.

\section{Related Work}

\paragraph{Type Systems for Rewrite Systems}
Term-rewriting systems \cite{dershowitz1985computing} have been shown useful in various applications such as program transformation, languages semantics and computations in theorem-proving systems. The rewrite rules in classical term-rewriting are terminating and confluent, which is the not the case for term-rewriting systems for program transformation with user defined strategies. In strategy languages such as ELAN \cite{borovansky1996elan}, Statego \cite{visser1998building, martin2008stratego} and TL \cite{winter2004special}, this issue is addressed by using composition operators to control reduction sequences, which is adapted in the design and implementation of \elevate~\cite{bastianhagedorn2020achieving}. Basic type systems and formalization for strategic term rewriting have been presented \cite{lammel2003typed, kaiser2009an} covering generic traversal. Our contribution advances previous work by providing more advanced types modeled using row polymorphism for rewriting programs.

\paragraph{Row Polymorphic Languages}
Introduced by R\'{e}my \cite{remy1989type} and Wand \cite{wand1991type}, row polymorphism is a parametric polymorphism allows representing an extensible structure in types as a row, usually a sequence of label-type pairs (while other notions exist \cite{morris19row}), which can be used as the basis for (polymorphic) variants and records. Row polymorphism is as expressive as structural subtyping, but works smoothly with Hindley-Milner style type inference, and it has many applications in modeling type systems. The structurally polymorphic types in OCaml are modeled using row polymorphism with guarantees on type soundness \cite{garrigue2015certified}. In Links \cite{daniel2016liberating}, row polymorphism is used for the implementation of algebraic effects and effect handlers, providing modular abstraction for effectful computation. More recently, \cite{morris19row} introduces a general theory for existing row polymorphic type systems, which focuses on row concatenation and gives rise to the language ROSE as a flexible tool for programming with extensible data types.

The syntax of \elevate is inspired by \textbf{MLPolyR} \cite{blume06cases}, which uses rows to realize polymorphic variants and records. \textbf{MLPolyR} treats cases or pattern matching branches as first-class values and allows pattern matching expressions to be extended with new cases at any moment. However, in \elevate, we do not support this kind of extension to get convenient exhaustiveness checking. Fortunately, as \cite{blume06cases} points out, this design choice can have the same extensibility as their language if the typing rule for pattern matching reduces/refines the type of the matched term case by case, which is exactly what \elevate does (see Section \ref{sec:type}).

In the presence of polymorphic variants, the typing and exhaustiveness checking for pattern matching is brought into consideration. \elevate and many other languages \cite{morris19row, daniel2016liberating, gaster1998records, pottier04match} bypass or do not consider performing type inference and exhaustiveness checking for deeply nested patterns by only including shallow/simple patterns in the formalization, and bridge the gap (if exists) with syntactical transformation \cite{gaster1998records}. Certainly, there are also other works dealing with deep pattern matchings directly \cite{garrigue2004typing, castagna2016set}. From the practical aspect, the overall behavior of pattern matching in \elevate is very close to that in OCaml. Due to the local constraint property of OCaml \cite{garrigue2015certified}, \elevate sometimes can give more precise types if there are shared row variables.

\section{Conclusion}
In this paper, we have presented a row-polymorphic type system for strategy language.
We have presented its formal definition, its practical implementation, and a case study of ensuring the safe composition of program transformations.
Our type system guarantees that strategies that type check do not fail due to a missing case in pattern matching and do not contain dead branches.

We are keen to explore future applications of our type system for verifying the correctness of program transformations exploiting the proposition-as-types interpretation as well as synthesizing program transformations from type specifications.


\bibliography{references}

\clearpage

\appendix

\section{Type Soundness Proof}
\label{app:proof}

\begin{lemma}[Type Extension]
\label{lem:tExt}
If $\Delta;\Gamma \vdash e: t$, and $\Gamma'(x) = \Gamma(x)$ for all $x \in fv(e)$, and $\Gamma'$ is well-formed with respect to $\Delta$, then $\Delta;\Gamma' \vdash e: t$.
\end{lemma}

\begin{lemma}[Kind Extension]
\label{lem:kExt}
If $\Delta;\Gamma \vdash e: t$, and $\Delta'(\alpha) = \Delta(\alpha)$ for all $\alpha \in ftv(\Gamma) \cup ftv(t)$, then $\Delta';\Gamma \vdash e: t$.
\end{lemma}

\begin{lemma}[Type Substitution]
\label{lem:tSubst}
If $\Delta;\Gamma \vdash e: t$, and $\mathbb{S}$ is a substitution where $Reg(\mathbb{S}) \subseteq \Delta$, and $\Gamma[\mathbb{S}]$ and $t[\mathbb{S}]$ is well-formed with respect to $\Delta$, then $\Delta;\Gamma[\mathbb{S}] \vdash e: t[\mathbb{S}]$.
\end{lemma}

\begin{lemma}[Generalization]
\label{lem:gen}
If $\Delta;\Gamma, x: \sigma \vdash e: t$, and $\Delta \vdash \sigma' \instRel \sigma$, then $\Delta;\Gamma, x: \sigma' \vdash e: t$.
\end{lemma}

\begin{lemma}[Substitution]
\label{lem:subst}
If $\Delta;\Gamma, x: \some{\forall \ (\alpha_i: \kappa_i).}{}{i}{0}{n}t \vdash e: t'$, and $\Delta, \some{\alpha_i: \kappa_i}{}{i}{0}{n};\cdot \vdash v: t$, and $\some{\alpha_i}{}{i}{0}{n} \cap ftv(\Gamma) = \varnothing$, then $\Delta;\Gamma \vdash e[x \mapsto v]: t'$.
\end{lemma}

\begin{proof}
  By induction on the derivation of 

  $\Delta;\Gamma, x: \some{\forall \ (\alpha_i: \kappa_i).}{}{i}{0}{n}t \vdash e: t'$

  \begin{itemize}
    \item \textbf{Case} $e = x'$
    \begin{itemize}
      \item When $x' \neq x$, we have $\Delta;\Gamma \vdash x': t'$ by $\ruleName{T-Var}$, so $\Delta;\Gamma \vdash x'[x \mapsto v]: t'$.
      \item When $x' = x$, we have $\Delta \vdash \some{\forall \ (\alpha_i: \kappa_i).}{}{i}{0}{n}t \instRel t'$ and $\some{\alpha_i}{}{i}{0}{n} \cap ftv(t') = \varnothing$ by $\ruleName{T-Var}$ and the rules for $\instRel$, which means \[\exists \mathbb{S}, Dom(\mathbb{S}) = \some{\alpha_i}{}{i}{0}{n}, t[\mathbb{S}] = t'\], then we have $\Delta, \some{\alpha_i: \kappa_i}{}{i}{0}{n};\cdot \vdash v:t[\mathbb{S}]$ by Lemma \ref{lem:tSubst}, which simplifies to $\Delta, \some{\alpha_i: \kappa_i}{}{i}{0}{n};\cdot \vdash v:t'$, hence $\Delta, \some{\alpha_i: \kappa_i}{}{i}{0}{n};\Gamma \vdash x'[x \mapsto v]:t'$. Thus, we have $\Delta;\Gamma \vdash x'[x \mapsto v]:t'$ by assumptions and Lemma \ref{lem:kExt}.
    \end{itemize}
    \item \textbf{Case} $e = (\lambda \ x' = e_1)$
    
    We have $\Delta;\Gamma, x: \some{\forall \ (\alpha_i: \kappa_i).}{}{i}{0}{n}t, x': t_0 \vdash e_1: t_1$ and $t' = t_0 \to t_1$ by $\ruleName{T-Lam}$, and we construct a substitution $\mathbb{S} = \some{\alpha_i}{}{i}{0}{n} \mapsto \some{\alpha'_i}{}{i}{0}{n}$ where $\some{\alpha'_i}{}{i}{0}{n}$ are distinct from all existing type variables. Let $\sigma = \some{\forall \ (\alpha_i: \kappa_i).}{}{i}{0}{n}t$, then we have 
      \begin{equation}
        \label{eq:substLam1}
        \Delta, \some{\alpha'_i: \kappa_i}{}{i}{0}{n};\Gamma, x': t_0, x: \sigma \vdash e_1: t_1
      \end{equation} by Lemma \ref{lem:kExt} and Lemma \ref{lem:tExt},
      \begin{equation}
        \label{eq:substLam2}
      \Delta, \some{\alpha'_i: \kappa_i}{}{i}{0}{n};\Gamma, x': t_0[\mathbb{S}], x: \sigma \vdash e_1: t_1[\mathbb{S}]
      \end{equation} by Lemma \ref{lem:tSubst}.
      On the other hand, we have
      \begin{equation}
        \label{eq:substLam3}
        \Delta, \some{\alpha'_i: \kappa_i}{}{i}{0}{n},\some{\alpha_i: \kappa_i}{}{i}{0}{n};\cdot \vdash v: t
      \end{equation} by Lemma \ref{lem:kExt}, and 
      \begin{equation}
        \label{eq:substLam4}
        \some{\alpha_i}{}{i}{0}{n} \cap ftv(\Gamma, x': t_0[\mathbb{S}]) = \varnothing
      \end{equation} by the assumption about $\mathbb{S}$. Thus, \[\Delta, \some{\alpha'_i: \kappa_i}{}{i}{0}{n};\Gamma, x': t_0[\mathbb{S}] \vdash e_1[x \mapsto v]: t_1[\mathbb{S}]\] by the induction hypothesis with (\ref{eq:substLam2}), (\ref{eq:substLam3}) and (\ref{eq:substLam4}). $\mathbb{S}$ is bijective, so it can be inverted and we get \[\Delta, \some{\alpha'_i: \kappa_i}{}{i}{0}{n};\Gamma, x': t_0 \vdash e_1[x \mapsto v]: t_1\] by Lemma \ref{lem:tSubst}, and then \[\Delta;\Gamma, x': t_0 \vdash e_1[x \mapsto v]: t_1\] by Lemma \ref{lem:kExt}. Finally, we have \[\Delta;\Gamma \vdash (\lambda \ x' \Rightarrow e_1)[x \to v]: t'\] by $\ruleName{T-Lam}$.
    \item \textbf{Case} $e = \keyword{let} \ f = e_1 \ \keyword{in} \ e_2$
    
    Let $\sigma = \some{\forall \ (\alpha_i: \kappa_i).}{}{i}{0}{n}t$, we have
    \begin{equation}
      \label{eq:substLet1}
      \Delta, \some{\alpha'_j: \kappa'_j}{}{j}{0}{m};\Gamma, x: \sigma \vdash e_1: t_1
    \end{equation} where $\some{\alpha'_j}{}{j}{0}{m} = ftv(t_1) \setminus ftv(\Gamma, x: \sigma)$ by $\ruleName{T-Let}$. Let $\Delta' = \some{\alpha'_j: \kappa'_j}{}{j}{0}{m}$, then we have
    \begin{equation}
      \label{eq:substLet2}
      \Delta, \Delta', \some{\alpha_i: \kappa_i}{}{i}{0}{n};\cdot \vdash v: t
    \end{equation} by Lemma \ref{lem:kExt}, and we get
    \begin{equation}
      \label{eq:substLet3}
      \Delta, \Delta';\Gamma \vdash e_1[x \mapsto v]: t_1
    \end{equation} by the induction hypothesis with (\ref{eq:substLet1}) and (\ref{eq:substLet2}).
    On the other hand, let $\sigma'_f = \some{\forall \ (\alpha'_j: \kappa'_j).}{}{j}{0}{m}t_1$, we have
    \begin{equation}
      \label{eq:substLet4}
      \Delta;\Gamma, f: \sigma'_f, x: \sigma \vdash e_2: t'
    \end{equation} by $\ruleName{T-Let}$ and Lemma \ref{lem:tExt}, and
    \begin{equation}
      \label{eq:substLet5}
      \some{\alpha_i}{}{i}{0}{n} \cap ftv(\Gamma, f: \sigma'_f) = \varnothing
    \end{equation} by the assumption $\some{\alpha_i}{}{i}{0}{n} \cap ftv(\Gamma) = \varnothing$ and the algebra of sets. Thus, we get
    \begin{equation}
      \label{eq:substLet6}
      \Delta;\Gamma, f: \sigma'_f \vdash e_2[x \mapsto v]: t'
    \end{equation} by the induction hypothesis with (\ref{eq:substLet4}) and (\ref{eq:substLet5}).

    However, we cannot jump to the conclusion now because $\sigma'_f$ is generalized with respect to $\Gamma, x: \sigma$ instead of $\Gamma$. Let $\some{\dot{\alpha}_p}{}{p}{0}{u} = ftv(t_1) \setminus ftv(\Gamma)$, and we know that $\some{\alpha'_j}{}{j}{0}{m} \subseteq \some{\dot{\alpha}_p}{}{p}{0}{u}$, so there exists $\dot{\Delta}$ and $\some{\kappa''_p}{}{p}{0}{u}$ such that\[\left(\Delta, \Delta'\right) = \left(\dot{\Delta}, \some{\dot{\alpha}_p: \kappa''_p}{}{p}{0}{u}\right)\]. Then we construct a substitution $\mathbb{S} = \some{\dot{\alpha}_p}{}{p}{0}{u} \mapsto \some{\alpha''_p}{}{p}{0}{u}$ where $\some{\alpha''_p}{}{p}{0}{u}$ are distinct from all existing type variables. Let $\Delta'' = \some{\alpha''_p: \kappa''_p}{}{p}{0}{u}$. On one hand, we have \[\Delta, \Delta', \Delta'';\Gamma \vdash e_1[x \mapsto v]: t_1[\mathbb{S}]\] by (\ref{eq:substLet3}), Lemma \ref{lem:kExt} and Lemma \ref{lem:tSubst}, then we safely remove $\Delta'$ and get
    \begin{equation}
      \label{eq:substLet7}
      \Delta, \Delta'';\Gamma \vdash e_1[x \mapsto v]: t_1[\mathbb{S}]
    \end{equation}
    by Lemma \ref{lem:kExt}. On the other hand, let \[\sigma''_f = \some{\forall \ (\alpha''_p: \kappa''_p).}{}{p}{0}{u}t_1[\mathbb{S}]\], we have 
    \begin{equation}
      \label{eq:substLet8}
      \Delta;\Gamma, f: \sigma''_f \vdash e_2[x \mapsto v]: t'
    \end{equation} by (\ref{eq:substLet6}) and Lemma \ref{lem:gen}. Finally, we get \[\Delta;\Gamma \vdash (\keyword{let} \ f = e_1 \ \keyword{in} \ e_2)[x \mapsto v]: t'\] by (\ref{eq:substLet7}), (\ref{eq:substLet8}) and $\ruleName{T-Let}$.
    \item \textbf{Case} $e = \keyword{match} \ e_1 \ \keyword{with} \ \langle
    l \ x_1 \Rightarrow rhs_1 \mid x_2 \Rightarrow rhs_2 \rangle$
    
    Let $\sigma = \some{\forall \ (\alpha_i: \kappa_i).}{}{i}{0}{n}t$, we have
    \begin{equation}
      \label{eq:substMatch1}
      \Delta, \some{\alpha'_j: \kappa'_j}{}{j}{0}{m};\Gamma, x: \sigma \vdash e_1: \langle l: t_1 \mid \rho \rangle
    \end{equation} where $\some{\alpha'_j}{}{j}{0}{m} = \some{\alpha'_p}{}{p}{0}{u} \cup \some{\alpha'_q}{}{q}{0}{v} = (ftv(t_1) \setminus ftv(\Gamma, x: \sigma)) \cup (ftv(\langle \rho \rangle) \setminus ftv(\Gamma, x: \sigma))$ by $\ruleName{T-Match}$. Let $\Delta' = \some{\alpha'_j: \kappa'_j}{}{j}{0}{m}$, then we have
    \begin{equation}
      \label{eq:substMatch2}
      \Delta, \Delta', \some{\alpha_i: \kappa_i}{}{i}{0}{n};\cdot \vdash v: t
    \end{equation} by Lemma \ref{lem:kExt}, and we get
    \begin{equation}
      \label{eq:substMatch3}
      \Delta, \Delta';\Gamma \vdash e_1[x \mapsto v]: \langle l: t_1 \mid \rho \rangle
    \end{equation} by the induction hypothesis with (\ref{eq:substMatch1}) and (\ref{eq:substMatch2}).
    On the other hand, let $\sigma'_{x_1} = \some{\forall \ (\alpha'_p: \kappa'_p).}{}{p}{0}{u}t_1$ and $\sigma'_{x_2} = \some{\forall \ (\alpha'_q: \kappa'_q).}{}{q}{0}{v}\langle \rho \rangle$, we have
    \begin{equation}
      \label{eq:substMatch4}
      \Delta;\Gamma, x_1: \sigma'_{x_1}, x: \sigma \vdash rhs_1: t'
    \end{equation} and
    \begin{equation}
      \label{eq:substMatch5}
      \Delta;\Gamma, x_2: \sigma'_{x_2}, x: \sigma \vdash rhs_2: t'
    \end{equation} by $\ruleName{T-Let}$ and Lemma \ref{lem:tExt}. Thus, we get
    \begin{equation}
      \label{eq:substMatch6}
      \Delta;\Gamma, x_1: \sigma'_{x_1} \vdash rhs_1[x \mapsto v]: t'
    \end{equation} and
    \begin{equation}
      \label{eq:substMatch7}
      \Delta;\Gamma, x_2: \sigma'_{x_2} \vdash rhs_2[x \mapsto v]: t'
    \end{equation} by the induction hypothesis with (\ref{eq:substMatch4}) and (\ref{eq:substMatch5}).

    Still, more proof steps are required because $\sigma'_{x_1}$ and $\sigma'_{x_2}$ are not generalized with respect to $\Gamma, x: \sigma$, but the same proof technique for $e = \keyword{let} \ f = e_1 \ \keyword{in} \ e_2$ can be used here, so part of the proof is omitted. Let $\mathbb{S}$ and $\Delta''$ be the newly constructed substitution and kinding enviroment, and $\sigma''_{x_1}$ and $\sigma''_{x_2}$ be the more generalized type schemes, we have
    \begin{equation}
      \label{eq:substMatch8}
      \Delta, \Delta'';\Gamma \vdash e_1[x \mapsto v]: \langle l: t_1 \mid \rho \rangle [\mathbb{S}]
    \end{equation} and
    \begin{equation}
      \label{eq:substMatch9}
      \Delta;\Gamma, x_1: \sigma''_{x_1} \vdash rhs_1[x \mapsto v]: t'
    \end{equation} and
    \begin{equation}
      \label{eq:substMatch10}
      \Delta;\Gamma, x_2: \sigma''_{x_2} \vdash rhs_2[x \mapsto v]: t'
    \end{equation}
    . Finally, we get \[
      \begin{aligned}
      &\Delta;\Gamma \vdash (\keyword{match} \ e \ \keyword{with} \\
      &\qquad \qquad \langle l \ x_1 \Rightarrow rhs_1 \mid x_2 \Rightarrow rhs_2 \rangle)[x \mapsto v] : t'
      \end{aligned}
    \] by $\ruleName{T-Match}$ with (\ref{eq:substMatch8}), (\ref{eq:substMatch9}) and (\ref{eq:substMatch10})

    \item For the rest of the cases, they can be routinely proven by applying the induction hypothesis. The proof for $e = fun \ arg$ is given here as an example: it follows from $\ruleName{T-App}$ and the induction hypothesis that \[\Delta;\Gamma \vdash fun[x \mapsto v]: t_1 \to t'\]\[\Delta;\Gamma \vdash arg[x \mapsto v]: t_1\], hence $\Delta;\Gamma \vdash fun \ arg[x \mapsto v]: t'$ by $\ruleName{T-App}$.
  \end{itemize}
\end{proof}

\begin{lemma}[Subject Reduction]
If $\Delta;\cdot \vdash e_1: t$, and $e_1 \evalRel e_2$, then $\Delta;\cdot \vdash e_2: t$.
\end{lemma}

\begin{proof}
  By case analysis on the reduction $e_1 \evalRel e_2$.
  \begin{itemize}
    \item \textbf{Case} $(\lambda \ x = e) \ v \evalRel e[x \mapsto v]$
    
    We have $\Delta;\cdot \vdash (\lambda \ x = e): t_1 \to t$ and $\Delta;\cdot \vdash v: t_1$ by $\ruleName{T-App}$, and then we get $\Delta; x: t_1 \vdash e: t$ by $\ruleName{T-Lam}$, and finally $\Delta;\cdot \vdash e[x \mapsto v]: t$ by Lemma \ref{lem:subst}.

    \item \textbf{Case} $\keyword{let} \ f = v \ \keyword{in} \ e_2 \evalRel e_2[f \mapsto v]$
    
    We have $\Delta, \some{\alpha_i: \kappa_i}{}{i}{0}{n};\cdot \vdash v: t_1$ where $\some{\alpha_i}{}{i}{0}{n} = ftv(t_1)$, and $\Delta;f: \some{\forall (\alpha_i: \kappa_i).}{}{i}{0}{n} t_1 \vdash e_2: t$ by $\ruleName{T-Let}$, and then $\Delta;\cdot \vdash e_2[f \mapsto v]: t$ by Lemma \ref{lem:subst}.

    \item \textbf{Case} \[\begin{aligned}
      &\keyword{match} \ l \ v \ \keyword{with} \ \langle l \ x_1 \Rightarrow e_1 \mid x_2 \Rightarrow e_2\rangle \\
      &\qquad \evalRel e_1[x_1 \mapsto v] 
      \end{aligned}\]

    We have $\Delta, \some{\alpha_i: \kappa_i}{}{i}{0}{n};\cdot \vdash l \ v : \langle l: t_1 \mid \rho \rangle$ where $\some{\alpha_i}{}{i}{0}{n} = \some{\alpha_p}{}{p}{0}{u} \cup \some{\alpha_q}{}{q}{0}{v} = ftv(t_1) \cup ftv(\langle \rho \rangle)$, and $\Delta; x_1: \some{\forall (\alpha_p: \kappa_p).}{}{p}{0}{u} t_1 \vdash e_1: t$ by $\ruleName{T-Match}$, and then we get $\Delta, \some{\alpha_p: \kappa_p}{}{p}{0}{u};\cdot \vdash v: t_1$ by $\ruleName{T-Label}$ and Lemma \ref{lem:kExt}, and finally $\Delta;\cdot \vdash e_1[x_1 \mapsto v]: t$ by Lemma \ref{lem:subst}.

    \item \textbf{Case} \[\begin{aligned}
      &\keyword{match} \ l \ v \ \keyword{with} \ \langle l' \ x_1 \Rightarrow e_1 \mid x_2 \Rightarrow e_2\rangle \\
      &\qquad \evalRel e_2[x_2 \mapsto l \ v] 
     \end{aligned}\]

    We have $\Delta, \some{\alpha_i: \kappa_i}{}{i}{0}{n};\cdot \vdash l \ v : \langle l': t_1 \mid \rho \rangle$ where $\some{\alpha_i}{}{i}{0}{n} = \some{\alpha_p}{}{p}{0}{u} \cup \some{\alpha_q}{}{q}{0}{v} = ftv(t_1) \cup ftv(\langle \rho \rangle)$, and $\Delta; x_2: \some{\forall (\alpha_q: \kappa_q).}{}{q}{0}{v} \langle \rho \rangle \vdash e_2: t$ by $\ruleName{T-Match}$, and then we get $\Delta, \some{\alpha_q: \kappa_q}{}{q}{0}{v};\cdot \vdash l \ v: \langle \rho \rangle$ where $\rho = (l: t' \mid \rho')$ and $\Delta, \some{\alpha_q: \kappa_q}{}{q}{0}{v};\cdot \vdash v: t'$ by $\ruleName{T-Label}$ and Lemma \ref{lem:kExt}, and finally $\Delta;\cdot \vdash e_2[x_2 \mapsto l \ v]: t$ by Lemma \ref{lem:subst}.

    \item For the rest of the cases, they can be routinely proven by applying the typing rules and their inversions. The proof for $\keyword{fix} \ v \evalRel v \ (\lambda \ x = \keyword{fix} \ v \ x)$ is given here as an example: we have $t = t_1 \to t_2$ and $\Delta;\cdot \vdash v: (t_1 \to t_2) \to t_1 \to t_2$ by $\ruleName{T-App}$ and $\ruleName{T-Fix}$, and then we have $\Delta;\cdot \vdash (\lambda \ x = \keyword{fix} \ v \ x): t_1 \to t_2$ by $\ruleName{T-App}$ and $\ruleName{T-Lam}$, and finally $\Delta;\cdot \vdash v \ (\lambda \ x = \keyword{fix} \ v \ x): t_1 \to t_2$ by $\ruleName{T-App}$.
  \end{itemize}
\end{proof}

\begin{lemma}[Canonical Forms]
  \label{lem:cf}
  If $v$ is a value,
  \begin{itemize}
    \item $v = l \ v'$ if $\Delta;\cdot \vdash v: \langle l: t' \mid \rho \rangle$.
    \item $v = \{\many{l_i: v_i}{\mid}{i}{0}{n}\}$ if $\Delta;\cdot \vdash v: \{\many{l_i: t_i}{\mid}{i}{0}{n}\}$.
    \item $v = (\lambda \ x = e)$ or $v = \keyword{fix}$ if $\Delta;\cdot \vdash v: t_1 \to t_2$.
  \end{itemize}
\end{lemma}

\begin{lemma}[Progress]
  If $\Delta;\cdot \vdash e: t$, then either $e$ is a value, or there exists an $\hat{e}$ such that $e \stepRel \hat{e}$.
\end{lemma}

\begin{proof}
  This lemma can be restated as: if $\Delta;\cdot \vdash e: t$, then either $e$ is a value, or there exist an evaluation context $E$ and two expressions $e'$ and $e''$ such that $e = E[e']$ and $e' \evalRel e''$.
  
  By induction on the derivation of $\Delta;\cdot \vdash e: t$.
  \begin{itemize}
    \item \textbf{Case} $e = x$ 
    
    Impossible.
    
    \item \textbf{Case} $e = (\lambda \ x = e_1)$
    
    $e$ is already a value.

    \item \textbf{Case} $e = fun \ arg$
    
    We have
    \begin{equation}
    \label{eq:progApp1}
      \Delta;\cdot \vdash fun: t_1 \to t
    \end{equation} and 
    \begin{equation}
      \label{eq:progApp2}
        \Delta;\cdot \vdash arg: t_1
    \end{equation} by $\ruleName{T-App}$. By applying the induction hypothesis on (\ref{eq:progApp1}), we get:
    \begin{itemize}
      \item If $fun = E_{fun}[e']$, and $e' \evalRel e''$, then there exists $E = E_{fun} \ arg$ such that $e = E[e']$.
      \item If $fun$ is a value, by applying the induction hypothesis on (\ref{eq:progApp2}), we get:
      \begin{itemize}
        \item If $arg = E_{arg}[e']$, and $e' \evalRel e''$, then there exists $E = fun \ E_{arg}$ such that $e = E[e']$.
        \item If $arg$ is a value, and $fun$ can be either $(\lambda \ x = e_1)$ or $\keyword{fix}$ by Lemma \ref{lem:cf}, then there exists a reduction with $E = []$ by $\ruleName{ST-App}$ and $\ruleName{ST-Fix}$, respectively.
      \end{itemize}
    \end{itemize}

    \item \textbf{Case} $e = \keyword{let} \ f = e_1 \ \keyword{in} \ e_2$
    
    We have
    \begin{equation}
      \label{eq:progLet}
      \Delta, \some{\alpha_i: \kappa_i}{}{i}{0}{n};\cdot \vdash e_1: t_1
    \end{equation} where $\some{\alpha_i}{}{i}{0}{n} = ftv(t_1)$ by $\ruleName{T-Let}$. By applying the induction hypothesis on (\ref{eq:progLet}), we get:
    \begin{itemize}
      \item If $e_1 = E_1[e']$, and $e' \evalRel e''$, then there exists $E = \keyword{let} \ f = E_1 \ \keyword{in} \ e_2$ such that $e = E[e']$.
      \item If $e_1$ is a value, there exists a reduction with $E = []$ by $\ruleName{ST-Let}$.
    \end{itemize}

    \item \textbf{Case} $e = \keyword{fix}$
    
    $e$ is already a value.

    \item \textbf{Case} $e = l \ e_1$
    
    We have
    \begin{equation}
    \label{eq:progLabel}
      \Delta;\cdot \vdash e_1: t_1
    \end{equation} by $\ruleName{T-Label}$. By applying the induction hypothesis on (\ref{eq:progLabel}), we get:
    \begin{itemize}
      \item If $e_1 = E_1[e']$, and $e' \evalRel e''$, then there exists $E = l \ E_1$ such that $e = E[e']$.
      \item If $e_1$ is a value, then $e$ is a value.
    \end{itemize}

    \item \textbf{Case} $e = \keyword{match} \ e \ \keyword{with} \ \langle\rangle$
    
    We have
    \begin{equation}
    \label{eq:progVoid}
      \Delta;\cdot \vdash e_1: \langle\cdot\rangle
    \end{equation} by $\ruleName{T-Void}$. By applying the induction hypothesis on (\ref{eq:progVoid}), we get:
    \begin{itemize}
      \item If $e_1 = E_1[e']$, and $e' \evalRel e''$, then there exists $E = \keyword{match} \ E_1 \ \keyword{with} \ \langle\rangle$ such that $e = E[e']$.
      \item It is impossible for $e_1$ to be a value. 
    \end{itemize}

    \item \textbf{Case} $e = \keyword{match} \ e_1 \ \keyword{with} \ \langle
    l \ x_1 \Rightarrow rhs_1 \mid x_2 \Rightarrow rhs_2
    \rangle$

    We have
    \begin{equation}
      \label{eq:progMatch}
      \Delta, \some{\alpha_i: \kappa_i}{}{i}{0}{n};\cdot \vdash e_1: \langle l: t_1 \mid \rho \rangle
    \end{equation} where $\some{\alpha_i}{}{i}{0}{n} = ftv(t_1) \cup ftv(\langle \rho \rangle)$ by $\ruleName{T-Let}$. By applying the induction hypothesis on (\ref{eq:progMatch}), we get:
    \begin{itemize}
      \item If $e_1 = E_1[e']$, and $e' \evalRel e''$, then there exists $E = \keyword{match} \ E_1 \ \keyword{with} \ \langle
      l \ x_1 \Rightarrow rhs_1 \mid x_2 \Rightarrow rhs_2
      \rangle$ such that $e = E[e']$.
      \item If $e_1$ is a value, it must be $(l' \ v)$ by Lemma \ref{lem:cf}, where $l'$ may or may not equal $l$, and then there exists a reduction with $E = []$ by $\ruleName{ST-Match-Match}$ and $\ruleName{ST-Match-Skip}$, respectively.
    \end{itemize}

    \item The rest of the cases can be routinely proven.

  \end{itemize}
\end{proof}

\clearpage

\end{document}